\documentclass[11pt, a4paper, logo, copyright, nonumbering]{map}
\usepackage{CJKutf8}
\usepackage{xargs} 
\usepackage{todonotes}
\usepackage{amsmath}
\usepackage{dsfont}

\usepackage{longtable}

\usepackage{svg}
\usepackage{fontawesome}
\usepackage{array}
\usepackage{tabularx}
\usepackage{latexsym}
\usepackage{graphicx}
\usepackage[utf8]{inputenc} 
\usepackage[utf8]{inputenc} 
\usepackage{amssymb} 
\usepackage{url}            
\usepackage{amsfonts}       
\usepackage{nicefrac}       
\usepackage{microtype}      
\usepackage{xspace}

\usepackage{listings}
\usepackage{makecell}
\usepackage{inconsolata}
\usepackage{multicol}
\usepackage{amstext}

\definecolor{darkblue}{RGB}{84, 112, 198}

\definecolor{lightblue}{rgb}{0.85, 0.95, 1.0}    
\definecolor{lightgreen}{rgb}{0.90, 1.0, 0.90}    
\definecolor{lightorange}{rgb}{1.0, 0.95, 0.85}   
\definecolor{lightpurple}{rgb}{0.95, 0.90, 1.0}   
\definecolor{lightgray}{rgb}{0.97, 0.97, 0.97}    

\usepackage{tikz}
\usetikzlibrary{calc}
\definecolor{battery-empty}{rgb}{0.9, 0.9, 0.9}
\newcommand{\difficultybar}[1]{%
  \begin{tikzpicture}[baseline, scale=0.5, every node/.style={scale=0.8}]
    \foreach \i in {1,2,3,4,5} {
      \ifnum\i>#1
        \draw[fill=battery-empty] (\i*0.5-0.5, 0) rectangle (\i*0.5, 0.25);
      \else
        \pgfmathsetmacro{\colorlevel}{80 - 12*(\i)} 
        \edef\x{\noexpand\draw[fill=blue!\colorlevel!white, opacity=0.9] (\i*0.5-0.5, 0) rectangle (\i*0.5, 0.25);}
        \x
        \draw[blue!50!black] (\i*0.5-0.5, 0) rectangle (\i*0.5, 0.25);
      \fi
    }
    \fill[battery-empty!70] (2.5, 0.08) rectangle (2.6, 0.17);
    \draw[battery-empty!70!black] (2.5, 0.08) rectangle (2.6, 0.17);
  \end{tikzpicture}%
}
\renewcommand{\arraystretch}{0.96}

\definecolor{hidden-draw}{RGB}{20,68,106}
\definecolor{hidden-pink}{RGB}{255,245,247}
\usepackage[edges]{forest}

\usepackage{bm}

\usepackage{natbib}
\usepackage{CJKutf8}
\usepackage{xargs}
\usepackage{todonotes}
\usepackage{multirow}
\usepackage{cleveref}
\usepackage{amsmath}
\usepackage{dsfont}
\usepackage{subcaption}
\usepackage{url}
\usepackage{svg}
\usepackage{fontawesome}
\usepackage{xcolor}
\usepackage{float}

\usepackage[utf8]{inputenc}
\usepackage{booktabs}
\usepackage{graphicx}
\usepackage{array}
\usepackage{tabularx}
\usepackage{xcolor,colortbl}  
\definecolor{boxcolor}{HTML}{d92523} 
\definecolor{bulbcolor}{HTML}{e3b87f} 
\usepackage{url}
\usepackage{ragged2e}  
\usepackage{makecell} 

\usepackage{hyperref}       
\usepackage{url}            
\usepackage{booktabs}       
\usepackage{amsfonts}       
\usepackage{nicefrac}       
\usepackage{microtype}      
\usepackage{xcolor}         
\usepackage{graphicx}
\usepackage{longtable}
\usepackage{array}
\usepackage{pbox}
\usepackage{amsmath}
\usepackage{amssymb}
\usepackage{tabularx}
\usepackage{multirow}
\usepackage{wrapfig}
\usepackage{pifont}
\usepackage{algorithm}
\usepackage{algorithmic}
\usepackage{lipsum}
\usepackage{subcaption}
\usepackage{enumitem}
\usepackage{tikz}
\usepackage{xstring}

\usepackage{color-edits}

\usepackage{booktabs}
\usepackage{multirow}
\usepackage[table]{xcolor}
\usepackage{tabularx}
\usepackage{graphicx}
\usepackage{hyperref}

\usepackage{parskip} 

\usepackage{threeparttable}

\newcommand{\modelname}{InCoder-32B} 
\definecolor{rliableolive}{HTML}{BBCC33}
\definecolor{rliableblue}{HTML}{77AADD}
\definecolor{rliablered}{HTML}{f63c44}

\definecolor{rliableolive}{HTML}{BBCC33}
\definecolor{rliableblue}{HTML}{77AADD}
\definecolor{rliablered}{HTML}{f63c44}

\usepackage[most,skins,theorems]{tcolorbox}
\tcbuselibrary{skins,breakable,fitting,hooks}  
\tcbset{
  aibox/.style={
    width=\linewidth,
    top=7pt,
    bottom=2pt,
    colback=rliablered!18!white,
    colframe=black,
    colbacktitle=black,
    enhanced,
    center,
    attach boxed title to top left={yshift=-0.1in,xshift=0.15in},
    boxed title style={boxrule=0pt,colframe=white,},
  }
}
\tcbset{
  aibox2/.style={
    width=\linewidth,
    top=7pt,
    bottom=2pt,
    colback=green!18!white,
    colframe=black,
    colbacktitle=black,
    enhanced,
    center,
    attach boxed title to top left={yshift=-0.1in,xshift=0.15in},
    boxed title style={boxrule=0pt,colframe=white,},
  }
}
\newtcolorbox{AIbox}[2][]{aibox,title=#2,#1}
\newtcolorbox{AIbox2}[2][]{aibox2,title=#2,#1}

\hypersetup{
    colorlinks=true,            
    linkcolor=blue,            
    filecolor=magenta,          
    urlcolor=cyan,             
    citecolor=purple,            
    pdftitle={Overleaf Example},
    pdfpagemode=FullScreen,
}

\definecolor{iquestblue}{HTML}{173C7F}
\definecolor{iquestazure}{HTML}{528FCC}

\renewcommand{\today}{}
\newcommandx{\info}[2][1=]{\todo[linecolor=red,backgroundcolor=red!25,bordercolor=red,#1]{#2}}

\title{
\vspace{-0.2in}
\centering \fontsize{15pt}{16pt}\selectfont
\modelname{}: Code Foundation Model for Industrial Scenarios
\vspace{-0.2in}
}

\author{
Jian Yang\textsuperscript{1}, 
Wei Zhang\textsuperscript{1}, 
Jiajun Wu\textsuperscript{1}, 
Junhang Cheng\textsuperscript{1}, 
Shawn Guo\textsuperscript{2},
Haowen Wang\textsuperscript{2},
Weicheng Gu\textsuperscript{1},
Yaxin Du\textsuperscript{3},
Joseph Li\textsuperscript{4},
Fanglin Xu\textsuperscript{2},
Yizhi Li\textsuperscript{5},
Lin Jing\textsuperscript{2},
Yuanbo Wang\textsuperscript{1},
Yuhan Gao\textsuperscript{1},
Ruihao Gong\textsuperscript{1},
Chuan Hao\textsuperscript{2},
Ran Tao\textsuperscript{2},
Aishan Liu\textsuperscript{1},
Tuney Zheng\textsuperscript{2}, 
Ganqu Cui\textsuperscript{6},
Zhoujun Li\textsuperscript{1},
Mingjie Tang\textsuperscript{7},
Chenghua Lin\textsuperscript{5},
Wayne Xin Zhao\textsuperscript{8}, 
Xianglong Liu\textsuperscript{1}, 
Ming Zhou\textsuperscript{9}, 
Bryan Dai\textsuperscript{2},
Weifeng Lv\textsuperscript{1}\\
\textsuperscript{1}Beihang University
\textsuperscript{2}IQuest Research
\textsuperscript{3}Shanghai Jiao Tong University
\textsuperscript{4}ELLIS
\textsuperscript{5}University of Manchester
\textsuperscript{6}Shanghai Artificial Intelligence Laboratory
\textsuperscript{7}Sichuan Univeristy
\textsuperscript{8}Gaoling School of Artificial Intelligence, Renmin University of China
\textsuperscript{9}Langboat
\\
\textsuperscript{$\dagger$}Corresponding Authors. Email: \texttt{\{jiayang\}@buaa.edu.cn}

{
\includegraphics[height=1em]{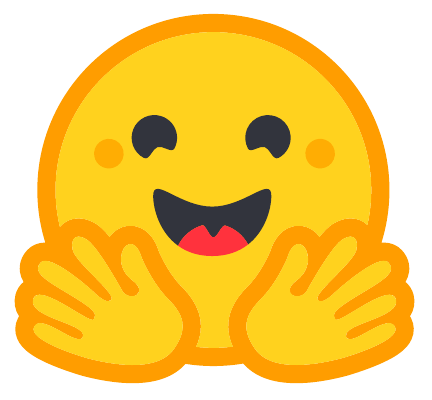}\;
HuggingFace: \url{https://huggingface.co/Multilingual-Multimodal-NLP/IndustrialCoder}
}\\

{
\includegraphics[height=1em]{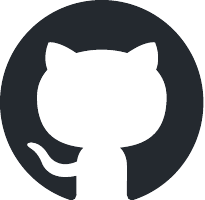}\;
GitHub: \url{https://github.com/CSJianYang/Industrial-Coder}
}
\vspace{-30pt}
}

\begin{abstract}
\vspace{-0.75em} 
Recent code large language models have achieved remarkable progress on general programming tasks. Nevertheless, their performance degrades significantly in industrial scenarios that require reasoning about hardware semantics, specialized language constructs, and strict resource constraints. To address these challenges, we introduce \modelname{} (Industrial-Coder-32B), the first 32B-parameter code foundation model unifying code intelligence across chip design, GPU kernel optimization, embedded systems, compiler optimization, and 3D modeling. By adopting an efficient architecture, we train InCoder-32B from scratch with general code pre-training, curated industrial code annealing, mid-training that progressively extends context from 8K to 128K tokens with synthetic industrial reasoning data, and post-training with execution-grounded verification. We conduct extensive evaluation on 14 mainstream general code benchmarks and 9 industrial benchmarks spanning 4 specialized domains. Results show InCoder-32B achieves highly competitive performance on general tasks while establishing strong open-source baselines across industrial domains.
\end{abstract}

\begin{document}

\maketitle

\let\oldthefootnote\thefootnote

\begin{figure*}[h!]
    \centering
    \includegraphics[width=0.8\textwidth]{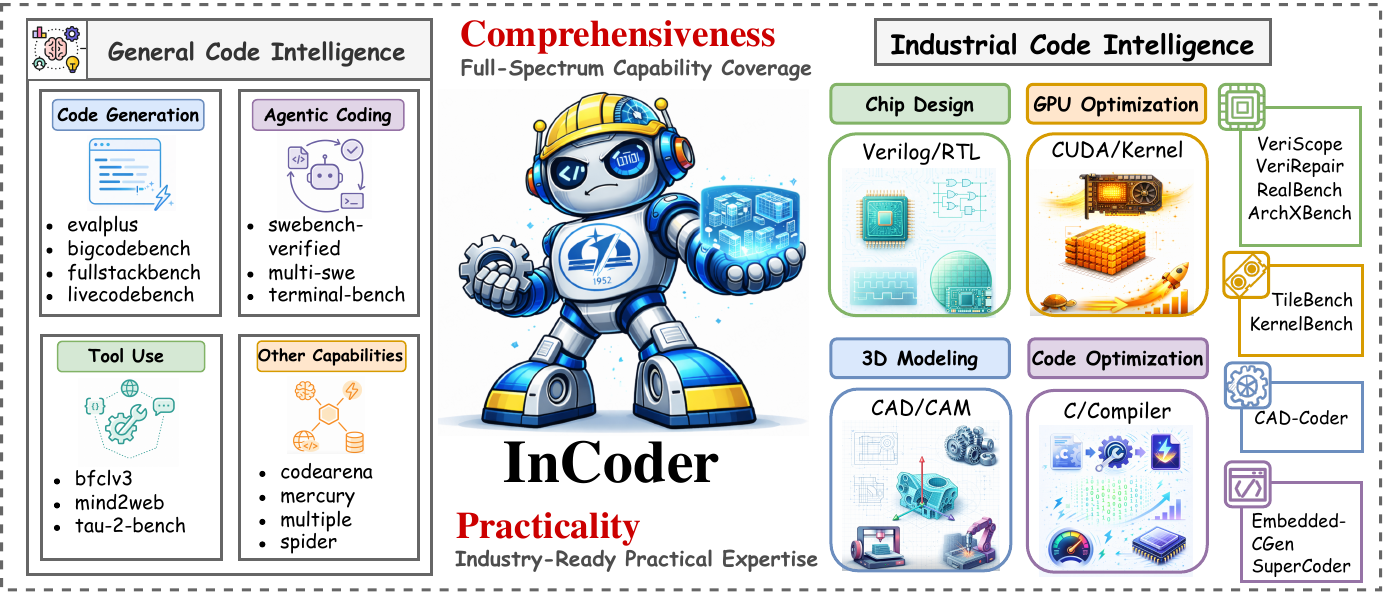}
    \vspace{-5pt}
    \caption{Scope of industrial code intelligence. \modelname~ aims to serve as a unified foundation model for (left) general software development and (right) industrial programming domains. The model supports a spectrum of capabilities, ranging from general coding tasks such as code generation, agentic development, and tool use to industrial workloads including chip design, GPU kernel optimization, CAD/CAM modeling, and compiler-level optimization.}
    \label{fig:tease_perf}
\end{figure*}

\newpage

\newpage
\begin{figure*}[t!]
    \centering
    \includegraphics[width=0.8\textwidth]{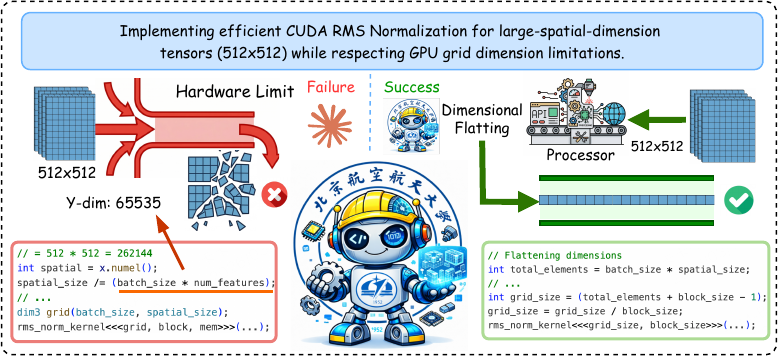}
    \caption{Comparison of CUDA grid configuration strategies for RMS Normalization on large spatial dimensions (512×512). (Left) Claude assigns spatial\_size (262,144) to gridDim.y, exceeding the CUDA hardware limit of 65,535, resulting in an invalid configuration argument runtime error. (Right) \modelname{} flattens all dimensions into a 1D grid, safely utilizing the gridDim.x limit and avoiding the hardware constraint violation.}
    \label{fig:intro_example}
\end{figure*}

\section{Introduction}

Code intelligence has witnessed substantial progress with the emergence of increasingly capable LLMs~\cite{yang2025codesurvey,iquestcoder_v1}. Recent model releases such as Qwen3.5~\cite{qwen3.5}, DeepSeek-V3.2~\cite{liu2025deepseekv32}, and Claude-4.6~\cite{claude46} have demonstrated strong performance across a wide range of programming tasks, with frontier models achieving gold-medal-level results in competitive programming~\cite{jain2024livecodebench}, software tasks~\cite{swebench,swe-bench-live,swebenchmultilingual}, and tool use tasks~\cite{tau2bench,servers25mcp}. These advances mark a turning point where LLMs have become genuinely capable assistants for everyday software engineering.

Much of this progress is driven by the abundance and diversity of publicly available code data. Repositories on GitHub, StackOverflow discussions, and open-source documentation provide rich supervision for training covering mainstream programming languages (PLs), frameworks, and development patterns. Yet a critical gap persists between general code intelligence and the demands of industrial software development. Scenarios such as CUDA kernel optimization~\cite{kernelbench}, Verilog hardware description~\cite{verilog_eval}, embedded firmware programming~\cite{firm_embedded_programming_llm}, and compiler optimization~\cite{compiler_optimization_llm} impose requirements that fundamentally differ from conventional software engineering with specialized language semantics, strict timing and resource constraints, reasoning about hardware behavior, and rigorous verification methodologies. Related benchmarks show that even the strongest code LLMs struggle on industrial tasks, with the best models achieving only 28.80\% call success rate of G and 41.57\% of T on Triton operator generation~\cite{li2025tritonbenchbenchmarkinglargelanguage} and 33.3\% accuracy of location generated Verilog code that passes simulation failing formal equivalence checking~\cite{jin2025realbench}.

To bridge this gap, we propose \textbf{\modelname}, \textbf{the first large language model purpose-built for industrial code intelligence}. With 32B parameters, \modelname{} is explicitly designed to tackle the unique challenges of industrial software development, including reasoning about hardware constraints, timing behavior, synthesis requirements, and low-level performance optimization, that existing code LLMs treat as out-of-distribution tasks. A single \modelname{} model serves chip design, GPU kernel optimization, embedded systems, compiler optimization, and 3D modeling, unifying these previously fragmented industrial domains for the first time. To achieve this, we adopt an efficient recurrent architecture and train \modelname{} through a three-stage \emph{Code-Flow} pipeline: (1)~\textbf{Pre-training \& Annealing} with curated industrial code data and automated verification; (2)~\textbf{Mid-training} that progressively extends context from 8K to 128K tokens with synthetic industrial reasoning data and agentic trajectories; and (3)~\textbf{Post-training} with execution-grounded verification. 

We conduct extensive evaluations on general and industrial code benchmarks and demonstrate that \modelname{} combines broad coding competence with specialized industrial capabilities. \modelname{} achieves 74.8\% on SWE-bench Verified, 49.14\% on LiveCodeBench, and 60.99\% on BFCL, competitive with leading models of comparable or larger scale. On industrial benchmarks, \modelname{} establishes the strongest open-source results across all evaluated domains, including chip design, GPU kernel optimization, embedded systems, compiler optimization, and 3D modeling.

Our contributions are:
\begin{itemize}
\item To the best of our knowledge, \modelname{} is the \textbf{first code LLM purpose-built for industrial code intelligence}, bridging the long-standing gap between academic code benchmarks and real-world industrial engineering domains such as chip design, GPU kernel optimization, embedded systems, and compiler engineering.

\item We assemble the most comprehensive industrial code evaluation to date, covering 14 general benchmarks and 9 industrial benchmarks across 4 specialized domains.

\item Through extensive ablations, we find that repository transition data outperforms static snapshots for planning, mid-training trajectory data improves robustness under distribution shift, and training with multi turn, feedback conditioned trajectories unlocks capabilities absent in standard instruction tuning.
\end{itemize}

\begin{figure*}[t!]
    \centering
    \includegraphics[width=\textwidth]{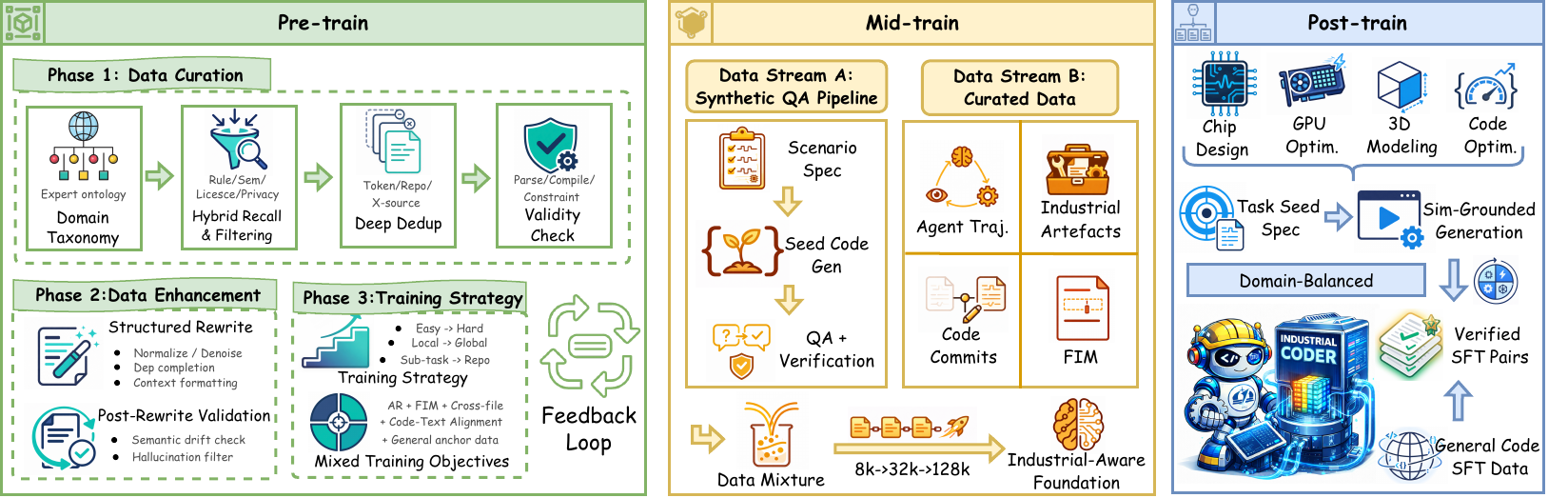}
    \caption{The three-stage training pipeline of \modelname{}. Pre-train performs data curation and enhancement, Mid-train constructs an industrial-aware foundation with progressive context scaling from 8K to 128K, and Post-train produces simulation-grounded SFT data across industrial domains.}
    \label{fig:pipeline}
\end{figure*}

\section{Scaling Industrial Data under Simulation Environments}
\label{sec:posttrain}

Industrial code differs from general software in that its correctness can only be established by running it in the same environment where it will ultimately be deployed. A Verilog module is validated through RTL simulation before it reaches silicon; a GPU kernel must execute on real hardware and produce numerically correct results; embedded firmware must boot on a microcontroller and interact correctly with its peripherals; and a CAD script must produce geometry that can be manufactured. To generate reliable post-training data for \modelname{}, we reconstruct these four classes of industrial environments in software, matching the toolchains and correctness criteria that engineers encounter in production.

\subsection{Chip Design}

In the semiconductor industry, a digital design progresses through an established flow: RTL authoring, behavioral simulation against testbenches, logic synthesis, and physical implementation. We reconstruct the first three stages using publicly available EDA tools. Icarus Verilog serves as the front end for behavioral simulation of Verilog designs. For IP cores written in SystemVerilog, we employ Verilator, which translates RTL into optimized C++ models and is the same simulator adopted by projects such as CHIPS Alliance and lowRISC. At the synthesis stage, Yosys maps RTL to a gate library, allowing us to verify synthesizability and extract area and timing estimates.

These three tools are composed into a single containerized image that mirrors the environment an RTL engineer works in: source files and testbenches go in, and compilation status, simulation results, and synthesis reports come out. By replicating this industrial flow rather than inventing a proxy, every training signal we extract is grounded in the same criteria that determine whether a design succeeds on real silicon.

\subsection{GPU Optimization}

GPU kernel development follows a distinct workflow: an engineer writes a kernel in CUDA or Triton, compiles it via the NVIDIA toolchain, launches it on a GPU, and validates both numerical correctness and performance. We replicate this workflow on NVIDIA A100 nodes.

For CUDA, we integrate the \texttt{nvcc} compiler through PyTorch's runtime compilation interface, matching the workflow used in libraries such as FlashAttention and xFormers where custom kernels are compiled and loaded at import time. For Triton, we rely on the official compiler stack: a Python function decorated with \texttt{@triton.jit} is compiled to GPU code at first invocation and cached for subsequent calls, the same path used in serving frameworks such as vLLM and SGLang.

The execution environment preserves the key characteristics of real deployment. Kernels launch on the same A100 hardware that production workloads target, memory is allocated through the standard CUDA allocator, and timing is measured via CUDA events. By building on the identical hardware and software stack that kernel engineers use, we ensure that signals obtained during data synthesis transfer directly to real deployment.

\subsection{3D Modeling}

In mechanical engineering, parametric CAD models are authored in scripting languages that drive a solid modeling kernel. The most widely adopted such kernel is OpenCascade, which supports Boolean operations, filleting, chamfering, extrusion, revolution, and lofting. CadQuery provides a Python interface to OpenCascade and has become the standard for programmatic CAD in the open hardware community.

We construct a modeling environment around CadQuery that reproduces the workflow a CAD engineer follows: a Python script defines geometric primitives, applies transformations, and exports the resulting solid to interchange formats such as STEP and STL. Generated scripts run against the same OpenCascade version used by production tools such as FreeCAD and KiCad, so code that passes our environment will also execute correctly in real CAD applications. Geometric fidelity is evaluated by tessellating the output solid and comparing it volumetrically against a reference, ensuring that the generated model is not merely syntactically valid but geometrically faithful to the specification.

\subsection{Code Optimization}

Code optimization in industry takes two forms: embedded systems programming, where code must run correctly on microcontrollers with specific peripheral hardware, and performance optimization, where the goal is to produce faster machine code. We construct a dedicated environment for each.

For embedded systems, we target the STM32F407, one of the most widely deployed ARM Cortex-M4 microcontrollers. The environment replicates the complete firmware toolchain: the \texttt{arm-none-eabi-gcc} cross compiler builds generated C code against CMSIS device headers and a linker script that maps the chip's memory layout. The compiled firmware is then loaded into the Renode simulator, which provides a virtual replica of the entire STM32F407 including GPIO ports, UART controllers, SPI and I2C buses, timers, ADC with DMA, and the interrupt controller. Each peripheral model reproduces the register layout and interrupt behavior specified in the reference manual, so that code running correctly in our environment will also run on physical hardware. This fidelity is critical because embedded bugs are often caused not by algorithmic errors but by incorrect register configuration or interrupt priority conflicts that only surface on real or faithfully emulated hardware.
 
For x86-64 assembly optimization, we replicate the standard compiler benchmarking workflow. Generated assembly is linked against a test harness and executed natively under controlled conditions: fixed CPU frequency, pinned core affinity, and repeated measurements. This mirrors the methodology used in LLVM and GCC regression suites, where the goal is to verify that an optimization is both correct and measurably faster.

The shared principle across all four environments is to replicate the toolchains and execution semantics that industrial engineers use rather than constructing simplified proxies. By building on the same simulators, compilers, and hardware that real deployments depend on, we ensure that training signals transfer directly to practice.

\begin{figure*}[t!]
    \centering
    \includegraphics[width=\textwidth]{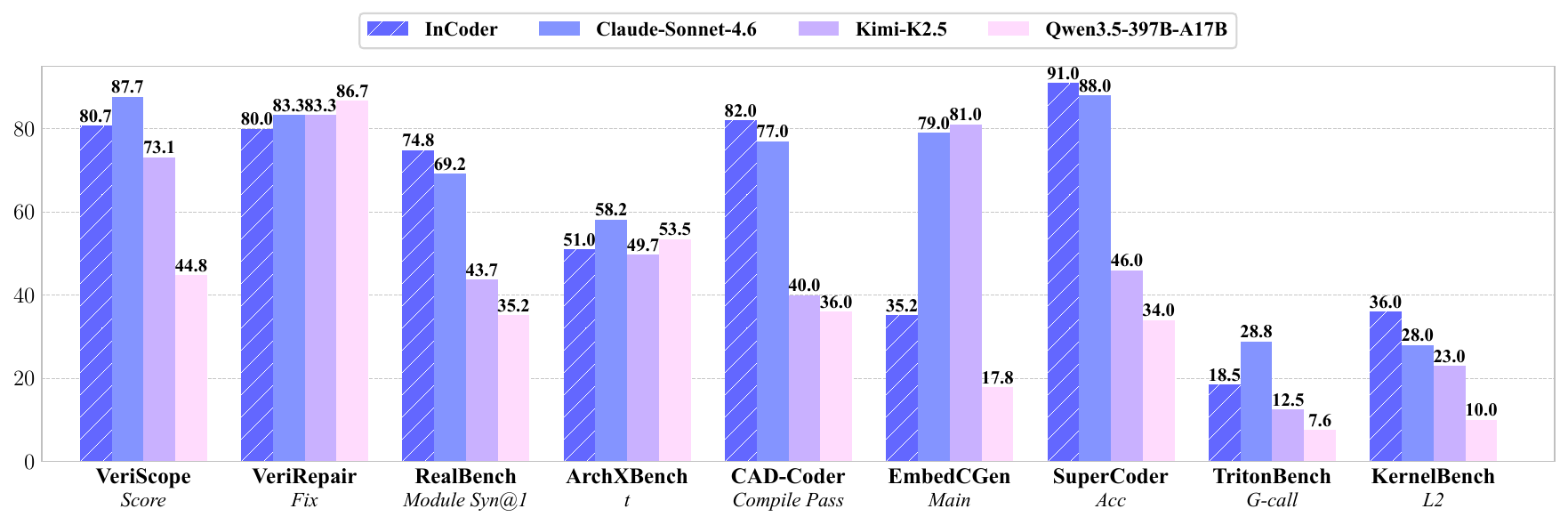}
    \caption{The performance of models on industrial code benchmarks.}
    \label{fig:benchmark_compare}
\end{figure*}

\section{Training Strategy}
\label{sec:pretraining}

Industrial hardware and system engineering spans diverse domains: digital circuit design (RTL/Verilog), GPU computing (Triton operators, CUDA kernels), systems programming (C/C++/Rust kernels), FPGA synthesis (HLS), CAD tool integration, and embedded systems—each with domain-specific challenges, timing constraints, resource budgets, and verification methodologies. While these domains showcase great coverage of industrial coding tasks, corresponding training corpora is lacking during the entire training stages. Detailed training procedures for pre-training, mid-training, and post-training are provided in Appendix~\ref{sec:pretraining_appendix}, \ref{sec:mid-training_appendix}, and \ref{sec:post-training_appendix}, respectively.

\subsection{Stage 1 Pre-Training}

\paragraph{Data Collection.}
We collect industrial codes from public repositories, technical literature, and domain-specific web data. Notably, we design a three-step recall strategy to increase the coverage of industrial codes we collect from public repositories. Additionally, we adopt OCR to collect high-quality code snippets and structured content from technical literature. Following the previous work~\cite{iquestcoder_v1}, we use the same model config for \modelname{}. See appendix for further details.

\paragraph{Data Cleaning and Refinement.}
We perform license filtering, personally identifiable information (PII) removal, and file-level validation, followed by deduplication at levels of exact hash matching, token-level near-duplicate detection~\cite{lozhkov2024starcoder2stackv2}, repository-level fork consolidation, and cross-source deduplication. We apply additional domain-specific checks before data refinement, where we normalize surface-level formatting and add structured annotations. All refined samples are verified through AST comparison and re-compilation to ensure correctness.

We train \modelname{} on 4,096 GPUs with autoregressive language modeling and fill-in-the-middle (FIM) completion~\cite{gong2025structureawarefillinthemiddlepretrainingcode, guo2024deepseek} using a standard decoder-only Transformer architecture. See appendix for more details.



\subsection{Stage 2 Mid-Training}
\label{sec:mid-training}

\subsubsection{Context Extension}
\label{subsec:training-strategy}

We extend model context length with a two-sub-stage strategy, increasingly extend from 8K tokens to 32K tokens, and then 128K tokens. While the first sub-stage focuses on file-level tasks, e.g. completing RTL modules, the latter sub-stage unlocks model's long-context capabilities, e.g. extended debugging sessions.

\subsubsection{Industrial Data Synthesis and Curation}
\label{subsec:data}

Our stage 2 pre-training data consist of synthetically generated industrial reasoning QAs, agent trajectories and code artifacts. Notably, our synthesized data leverage real-world development scenarios extensively that are normally underrepresented among public repositories.

\paragraph{Synthetic Industrial Code QA.}
Our synthesis pipeline operates in three steps designed to produce industrially grounded,
factually correct reasoning data:
\textit{(i) Industrial scenario specification} through consultation with practising hardware
and systems engineers;
\textit{(ii) Seed code generation} that reflects realistic hardware design patterns and domain-specific conventions;
\textit{(iii) QA pair synthesis with automated verification}.
The detailed synthesis pipeline and coverage analysis are provided in \autoref{app:synthetic-qa}.

\paragraph{Agent Trajectories.}
We include multi-step debugging and repair trajectories following the Thought-Action-Observation cycle~\cite{yang2024sweagentagentcomputerinterfacesenable}, capturing closed-loop reasoning with
tool feedback from hardware simulators, synthesis tools, C/C++ compilers, and formal verification engines. Curating such trajectories addresses the lack of operational context in standard code corpora.

\paragraph{Industrial Code Artifacts.}
We also include auxiliary artifacts that reflect the operational context
of professional hardware development: hardware testbenches (SystemVerilog/UVM), timing constraints (SDC),
synthesis scripts, GPU profiling traces, and memory sanitiser
logs~\cite{srivatsa2024surveyusinglargelanguage,alshahwan2024automatedunittestimprovement}.
These domain-specific artifacts expose the model to the full ecosystem of industrial hardware engineering,
compensating for their scarcity in public data.

\subsection{Stage3 Post-Training}

\paragraph{Data Construction}

General-purpose supervised fine-tuning (SFT) datasets~\cite{ouyang2022training, codealpaca} carry little signal for industrial coding tasks, especially when execution-based verifications can have a non-trivial impact. Therefore, we construct 2.5M samples directly from real-life industrial coding tasks grounded in execution. Finally, our tasks spanning across hardware design, GPU kernel development, systems programming, and embedded firmware.

\paragraph{Task Construction, Candidate Generation, and Verification}
Each task is decomposed into a structured instruction with a natural language requirement description, interface constraints (port lists, function signatures, API contracts), the target platform and toolchain, dependency configurations, and associated verification scripts. This normalization step produces a consistent instruction format for SFT.

Given an instruction, we generate a diverse set of candidate solutions through a group of complementary samples, such as template-based perturbation and cross-language migration, in order to boost the diversity of generated solutions. We validate generated solutions through execution. Notably, this verification is grounded in a real execution environment, i.e. where a real engineer use in production.

\paragraph{Feedback-Driven Repair}

For a solution that fail executions, our pipeline captures the entire feedback context, including compiler error messages, runtime logs, counterexample inputs, waveform differences, and profiling bottlenecks. We then append this feedback to the failed solution to generate a repaired solution. Note that the result is a closed-loop repair trajectory~\cite{ye2022selfapr, jiang2024ledex} including both the failed and the succeeded solution with execution feedback, which we also include in the SFT corpus in order to mimic a workflow of bug-fixing from an experienced engineer. 

\paragraph{Quality Filtering and Final Composition}
Finally, we filter SFT samples through executability, stability, and information density, from which we categorize samples into three kinds, i.e. \textit{direct solution}, \textit{defect repairs}, and \textit{performance and structural optimization samples}. Note that the last category refers to a correct solution improved with respect to efficiency, readability, or architectural quality. 

\definecolor{tablegray}{gray}{0.92}
\begin{table*}[t!]
    \centering
    \caption{Performance comparison on code generation tasks.}
    \label{tab:code_generation_1}
    \resizebox{1.0\textwidth}{!}{
    \begin{tabular}{lr|cccc|cc|c}
    \toprule
        \multirow{2}{*}{\textbf{Model}} & \multirow{2}{*}{\textbf{Size}} & \multicolumn{4}{c}{\textbf{EvalPlus}} & \multicolumn{2}{|c|}{\textbf{BigCodeBench}} & \multirow{2}{*}{\textbf{FullStackBench}} \\
        ~ & ~ & \textbf{HumanEval} & \textbf{HumanEval+} & \textbf{MBPP} & \textbf{MBPP+} & \textbf{Full} & \textbf{Hard} & ~  \\
        \midrule
        \multicolumn{9}{c}{\textbf{6B+ Models}} \\
        \midrule
        DeepSeek-Coder-V2-Lite-Instruct & 2.4/16B & 81.1 & 75.6 & 85.2 & 70.6 & 37.8 & 18.9 & 49.4 \\
        Qwen2.5-Coder-7B-Instruct & 7B & \textbf{87.2} & \textbf{81.7} & 84.7 & 72.2 & 37.8 & 13.5 & 42.2 \\
        Seed-Coder-8B-Instruct & 8B & 81.1 & 75.6 & 86.2 & 73.3 & 44.6 & \textbf{23.6} & \textbf{55.8} \\
        Qwen2.5-Coder-14B-Instruct & 14B & 62.8 & 59.8 & \textbf{88.6} & \textbf{77.2} & \textbf{47.0} & 6.1 & 53.1 \\
        \midrule
        \multicolumn{9}{c}{\textbf{30B+ Models}} \\
        \midrule
        Qwen3-Coder-30B-A3B-Instruct & 3.3/30.5B & 93.9 & 87.2 & 90.7 & 77.2 & 46.9 & 27.7 & 60.9 \\
        Deepseek-V3.2 & 37/671B & 93.9 & 88.4 & 93.4 & 77.2 & 48.1 & 27.0 & 64.9 \\
        Qwen2.5-Coder-32B-Instruct & 32B & 93.3 & 86.6 & 90.2 & 77.8 & 48.0 & 24.3 & 57.4 \\
        Qwen3-235B-A22B-Instruct-2507 & 22/235B & 96.3 & 91.5 & 92.3 & 77.8 & 47.4 & 25.7 & 62.7 \\
        Qwen3-235B-A22B-Thinking-2507 & 22/235B & \textbf{98.8} & \textbf{93.3} & 95.5 & 81.5 & 44.1 & 23.0 & - \\
        Qwen3-Coder-480B-A35B-Instruct & 35/480B & 97.6 & 92.7 & 94.2 & 80.2 & 49.4 & 27.7 & 66.4 \\
        Kimi-Dev-72B & 72B & 93.3 & 86.0 & 79.6 & 68.8 & 45.4 & \textbf{31.8} & 38.6 \\
        Kimi-K2-Instruct-0905 & 32B/1T & 94.5 & 89.6 & 91.8 & 74.1 & \textbf{49.8} & 30.4 & 63.5 \\
        Kimi-K2-Thinking & 32B/1T & 98.2 & 92.7 & \textbf{97.4} & \textbf{82.3} & 46.8 & 28.4 & - \\
        KAT-Dev & 32B & 90.9 & 86.6 & 89.4 & 76.2 & 46.2 & 25.7 & 58.8 \\
        KAT-Dev-72B-Exp & 72B & 88.4 & 81.7 & 85.2 & 69.3 & 48.3 & 26.4 & 52.9 \\
        GLM-4.7 & 32/355B & 87.2 & 79.9 & 90.5 & 75.7 & 45.7 & 26.4 & \textbf{70.2} \\
        \rowcolor{tablegray} \textbf{\modelname{}} & 32B & 94.5 & 89.6 & 91.8 & 78.3 & \textbf{49.8} & 31.1 & 57.1 \\
    \bottomrule
    \end{tabular}
    }
\end{table*}

\definecolor{tablegray}{gray}{0.92}
\begin{table*}[!h]
    \centering
    \caption{Combined performance on code reasoning (CruxEval, LiveCodeBench), code efficiency (Mercury), and Text2SQL (Bird, Spider) benchmarks.}
    \label{tab:code_reasoning_efficiency_sql}
    \resizebox{1.0\textwidth}{!}{
    \begin{tabular}{lr|cccc|cc|cc}
    \toprule
        \multirow{3}{*}{\textbf{Model}} & \multirow{3}{*}{\textbf{Size}} & \multicolumn{4}{c|}{\textbf{Code Reasoning}} & \multicolumn{2}{c|}{\textbf{Code Efficiency}} & \multicolumn{2}{c}{\textbf{Text2SQL}} \\
        \cmidrule(lr){3-6} \cmidrule(lr){7-8} \cmidrule(lr){9-10}
        ~ & ~ & \multicolumn{2}{c}{\textbf{CruxEval}} & \multicolumn{2}{c|}{\textbf{LiveCodeBench}} & \multicolumn{2}{c|}{\textbf{Mercury}} & \multirow{2}{*}{\textbf{Bird}} & \multirow{2}{*}{\textbf{Spider}} \\
        ~ & ~ & \textbf{Input-COT} & \textbf{Output-COT} & \textbf{V5} & \textbf{V6} & \textbf{Beyond@1} & \textbf{Pass@1} & ~ & ~ \\
        \midrule
        \multicolumn{10}{c}{\textbf{6B+ Models}} \\
        \midrule
        DeepSeek-Coder-V2-Lite-Instruct & 2.4/16B & 57.1 & 56.2 & 13.2 & 19.4 & 76.8 & 91.4 & 41.6 & 72.4 \\
        Qwen2.5-Coder-7B-Instruct & 7B & 66.9 & 66.0 & 14.4 & 18.9 & 69.9 & 84.8 & 53.1 & 79.8 \\
        Seed-Coder-8B-Instruct & 8B & 62.0 & 66.6 & 19.2 & 22.3 & \textbf{78.5} & \textbf{93.8} & 44.7 & 72.7 \\
        Qwen2.5-Coder-14B-Instruct & 14B & \textbf{75.6} & \textbf{79.2} & \textbf{22.8} & \textbf{24.6} & 76.7 & 88.3 & \textbf{59.1} & \textbf{81.3} \\
        \midrule
        \multicolumn{10}{c}{\textbf{30B+ Models}} \\
        \midrule
        Qwen3-Coder-30B-A3B-Instruct & 3.3/30.5B & 76.9 & 80.5 & 43.1 & 36.0 & 81.1 & 95.3 & 59.0 & 80.9 \\
        DeepSeek-v3.2 & 37/671B & 82.1 & \textbf{94.2} & - & 83.3 & \textbf{81.6} & \textbf{96.9} & 52.6 & 77.9 \\
        Qwen2.5-Coder-32B-Instruct & 32B & 78.8 & 84.0 & 30.5 & 27.4 & 79.1 & 96.1 & 62.1 & \textbf{83.9} \\
        Qwen3-235B-A22B-Instruct-2507 & 22/235B & 62.0 & 89.5 & 53.9 & 51.8 & 80.4 & \textbf{96.9} & \textbf{62.8} & 81.1 \\
        Qwen3-235B-A22B-Thinking-2507 & 22/235B & 15.2 & 46.9 & \textbf{80.2} & 74.1 & 61.2 & 70.3 & 35.2 & 42.6 \\
        Qwen3-Coder-480B-A35B-Instruct & 35/480B & 87.1 & 90.4 & 48.6 & 53.9 & 80.2 & 96.1 & 61.3 & 81.2 \\
        Kimi-Dev-72B & 72B & 33.0 & 64.2 & 46.1 & 40.0 & 59.1 & 69.5 & - & - \\
        Kimi-K2-Instruct-0905 & 32B/1T & 86.8 & 89.5 & 52.1 & 53.7 & 76.1 & 90.6 & 60.4 & 81.1 \\
        Kimi-K2-Thinking & 32B/1T & \textbf{92.2} & 86.2 & - & 83.1 & 73.0 & 85.2 & 40.6 & 49.6 \\
        KAT-Dev & 32B & 42.5 & 65.1 & 32.9 & 32.6 & 75.1 & 89.1 & 52.2 & 77.6 \\
        KAT-Dev-72B-Exp & 72B & 71.4 & 81.1 & 13.8 & 16.0 & 79.0 & 94.5 & 35.2 & 60.3 \\
        GLM-4.7 & 32/355B & 65.6 & 81.2 & - & \textbf{84.9} & 74.1 & 86.7 & 46.5 & 62.4 \\
        \rowcolor{tablegray} \textbf{\modelname{}} & 32B & 62.4 & 73.9 & 53.3 & 49.1 & 71.4 & 85.6 & 55.4 & 79.7 \\
    \bottomrule
    \end{tabular}
    }
\end{table*}

\begin{table*}[h]
    \centering
    \caption{Combined performance on agentic coding tasks (Terminal-Bench v1.0, Terminal-Bench v2.0, SWE-Verified) and general tool-use tasks (Mind2Web, BFCL V3, $\tau^2$-bench).}
    \label{tab:agentic_combined}
    \small
    \setlength{\tabcolsep}{3.2pt}
    \renewcommand{\arraystretch}{1.05}
    \resizebox{1.0\textwidth}{!}{
    \begin{tabular}{lr|ccc|ccccc}
    \toprule
        \multirow{3}{*}{\textbf{Model}} & \multirow{3}{*}{\textbf{Size}} &
        \multicolumn{3}{c|}{\textbf{Agentic Coding}} &
        \multicolumn{5}{c}{\textbf{General Tool Use}} \\
        \cmidrule(lr){3-5} \cmidrule(lr){6-10}
        & & \multicolumn{2}{c}{\textbf{Terminal-Bench}} & \multirow{2}{*}{\textbf{\shortstack{SWE-bench\\Verified}}}
        & \multirow{2}{*}{\textbf{Mind2Web}} & \multirow{2}{*}{\textbf{BFCL V3}} & \multicolumn{3}{c}{\textbf{$\tau^2$-bench}} \\
        & & \textbf{v1.0} & \textbf{v2.0} & & & & \textbf{Airline} & \textbf{Retail} & \textbf{Telecom} \\
    \midrule

        \multicolumn{10}{c}{\textbf{6B+ Models}} \\
    \midrule
        DeepSeek-Coder-V2-Lite-Instruct & 2.4/16B & 5.0 & 0.0 & -   & 26.7 & - & 3.5 & 12.0 & - \\
        Qwen2.5-Coder-7B-Instruct       & 7B & 6.3 & 0.0 & -   & 38.4 & 54.2 & - & - & - \\
        Seed-Coder-8B-Instruct          & 8B & 7.5 & \textbf{2.5} & -   & 38.2 & - & \textbf{4.3} & \textbf{32.0} & - \\
        Qwen2.5-Coder-14B-Instruct      & 14B & \textbf{8.8}  & 0.0  & -    & \textbf{42.7} & \textbf{59.9} & - & - & - \\
    \midrule
        \multicolumn{10}{c}{\textbf{30B+ Models}} \\
    \midrule
        Qwen3-Coder-30B-A3B-Instruct    & 3.3/30.5B & 23.8 & 23.8 & 51.9 & 36.1 & 63.4 & 42.0 & 25.4 & 25.4 \\
        DeepSeek-v3.2                   & 37/671B & 23.8 & \textbf{46.4} & 73.1 & 47.2 & 68.8 & 63.8 & 81.1 & \textbf{96.2} \\
        Qwen2.5-Coder-32B-Instruct      & 32B & 5.0  & 4.5  & -    & 32.5 & 62.3 & - & - & - \\
        Qwen3-235B-A22B-Instruct-2507   & 22/235B & 15.0 & 13.5 & 45.2 & 49.0 & 71.2 & 50.0 & 74.6 & 32.5 \\
        Qwen3-235B-A22B-Thinking-2507   & 22/235B & 8.8  & 3.4  & 44.6 & 43.2 & \textbf{71.9} & 58.0 & 71.9 & 45.6 \\
        Qwen3-Coder-480B-A35B-Instruct  & 35/480B & 37.5 & 23.6 & 67.0 & 54.0 & 68.7 & 60.0 & 77.5 & 65.8 \\
        Kimi-Dev-72B                    & 72B & -    & 2.3  & 60.4 & -    & 55.5 & 21.9 & 32.0 & 35.1 \\
        Kimi-K2-Instruct-0905           & 32B/1T & 44.5 & 27.8 & 69.2 & 53.4 & 70.3 & 56.5 & 70.6 & 65.8 \\
        Kimi-K2-Thinking                & 32B/1T & \textbf{47.1} & 33.7 & 71.3 & 55.7 & - & - & - & - \\
        KAT-Dev                         & 32B & 17.5 & 10.1 & 62.4 & 33.7 & 64.7 & 32.0 & 28.0 & 35.1 \\
        KAT-Dev-72B-Exp                 & 72B & 21.3 & 7.9  & 74.6 & -    & - & - & - & - \\
        GLM-4.7                         & 32/355B & 36.3 & 41.0 & 73.8 & 53.7 & 64.8 & 60.0 & 70.2 & 75.4 \\
        \rowcolor{tablegray}\textbf{\modelname{}}      & 32B & 35.0 & 22.5 & \textbf{74.8} & \textbf{55.8} & 61.0 & \textbf{70.0} & \textbf{85.1} & 86.8 \\
    \bottomrule
    \end{tabular}
    }
\end{table*}


\section{Evaluation}\label{sec:evaluation}
\subsection{Baselines}
We compare \modelname{} against a comprehensive set of large language models spanning both open-weight and proprietary systems, evaluating them across general-purpose code benchmarks and specialized industrial code domains.

For general-purpose code evaluation, our baselines include DeepSeek-Coder-V2-Lite-Instruct~\cite{deepseek2024coder} and DeepSeek-V3.2~\cite{liu2025deepseekv32}, the Qwen2.5-Coder series (7B, 14B, and 32B)~\cite{qwen25coder}, Qwen3-235B-A22B-Instruct and Qwen3-235B-A22B-Thinking~\cite{yang2025qwen3}, the Qwen3-Coder series (30B-A3B and 480B-A35B)~\cite{qwen3coder}, Seed-Coder-8B-Instruct~\cite{seedcoder} from ByteDance, Kimi-Dev-72B~\cite{kimi_dev}, Kimi-K2-Instruct and Kimi-K2-Thinking~\cite{team2025kimi2} from Moonshot AI, KAT-Dev and KAT-Dev-72B-Exp~\cite{katcoder} from Kuaishou, and GLM-4.7~\cite{glm47} from Zhipu AI.

For specialized industrial code evaluation, we evaluate DeepSeek-V3.2~\cite{liu2025deepseekv32}, the GLM series (GLM-5~\cite{zeng2026glm} and GLM-4.7~\cite{glm47}) from Zhipu AI, the Kimi family (Kimi-K2.5~\cite{team2026kimi25} and Kimi-K2 in both Instruct and Thinking variants~\cite{team2025kimi2}) from Moonshot AI, MiniMax-M2.5~\cite{minimax-m25}, the Qwen ecosystem comprising the general-purpose Qwen3.5 series ranging from 0.8B to 397B-A17B~\cite{qwen3.5}, Qwen3-Next~\cite{qwen3next}, and the code-specialized Qwen3-Coder series~\cite{qwen3coder,qwen_qwen3_coder_next_tech_report} from Alibaba, Seed-OSS-36B-Instruct~\cite{seed2025seed-oss} from ByteDance, and GPT-OSS (120B and 20B)~\cite{agarwal2025gpt} from OpenAI. For proprietary models, we include Claude-Sonnet-4.6~\cite{claude46} from Anthropic. These baselines collectively encompass dense and mixture-of-experts architectures across a wide parameter range, enabling a thorough investigation of current capability boundaries across both general code and industrial code domains.

\subsection{Benchmarks}

\subsubsection{General Code Evaluation}
We evaluate model performance across multiple dimensions: code generation using EvalPlus~\cite{evalplus} (HumanEval~\citep{chen2021codex} and MBPP~\citep{austin2021mbpp}), BigCodeBench~\cite{zhuo2024bigcodebench} for library-intensive tasks, and FullStackBench~\cite{liu2024fullstackbench} for full-stack scenarios; code reasoning with CRUXEval~\cite{gu2024cruxeval} testing bidirectional execution prediction (I2O and O2I) and LiveCodeBench~\cite{jain2024livecodebench} for competitive programming; code efficiency via Mercury~\cite{du2024mercurycodeefficiencybenchmark}, which jointly measures correctness and runtime performance; Text-to-SQL capabilities on Spider~\cite{2018spider} for schema linking and BIRD~\cite{2023bird} for value grounding; agentic coding tasks including Terminal-Bench~\cite{tbench2025} for terminal workflows, SWE-bench~\cite{jimenez2024swebench} for real-world patch generation, and SWE-bench Verified~\cite{swebenchverified} with human-curated instances; and general agentic tasks such as Mind2Web~\cite{deng2023mind2web} for web navigation, BFCL~\cite{patil2025bfcl} for multi-turn function calling across heterogeneous APIs, and $\tau$-bench~\cite{yao2024tau} with $\tau^2$-bench~\cite{barres2025tau2benchevaluatingconversationalagents} for policy-constrained conversational agents in shared environment.
 of
\subsection{Industrial Code Benchmarks}

As shown in \autoref{fig:benchmark_compare}, we also evaluate our model on industrial code tasks, i.e. tasks related to chip design, GPU kernel optimization, code optimization, and 3D modeling. These tasks differ from conventional software engineering in important ways: they require reasoning on hardware constraints, low-level performance trade-offs, and domain-specific correctness criteria.

\subsubsection{Chip Design}

\paragraph{VeriScope.}
We propose the Verilog generation benchmark comprising 568 problems across five difficulty levels with problems ranging from basic combinational logic, hierarchical module composition, system-level designs, to extreme challenges such as a dual-core out-of-order RISC-V SoC with cache coherence at L5. Each problem is evaluated through simulation: a code is scored 0/50/100 when it fails to compile, compiles but fails the test, or passes all unit tests.

\paragraph{RealBench.}
RealBench~\cite{jin2025realbench} targets production-grade IP-level Verilog generation rather than small algorithmic exercises. Built on four real-world open-source IP cores (AES encryption, SD card controller, and Hummingbirdv2 E203 CPU), it includes 60 module-level subtasks where sub-modules can fall back to golden implementations, and 4 system-level subtasks that require implementing the entire module hierarchy from scratch given only top-level specifications. Evaluation employs a layered verification pipeline: Syn@$k$ checks whether at least one of $k$ independently generated candidates compiles successfully, while Func@$k$ further verifies functional correctness via testbench simulation on compilable candidates, where $k$ denotes the number of independent samples. We report both metrics at $k \in \{1, 5\}$ for system-level and module-level tasks separately.

\paragraph{ArchXBench.}
ArchXBench~\cite{purini2025mlcad} provides 51 complex digital system designs across six difficulty levels with domains including cryptography, signal processing, image processing, and machine learning. Designs range from arithmetic circuits to complete subsystems like AES cores and streaming FFT/DCT pipelines. Each task includes a problem description, an interface specification, and a testbench. We generate five independent candidates per task and report results using an $(n, t)$ format, where $n$ denotes the number of syntactically correct candidates out of five and $t$ measures the percentage of testbench assertions passed by the best candidate among them.

\paragraph{VeriRepair.}
VeriRepair is our proposed Verilog error diagnosis and repair dataset, constructed by systematically injecting realistic bugs into correct implementations from VeriCoder-RTLCoder. The error taxonomy spans 4~major categories and 20~error types, including syntax errors, type and structural errors, timing and FSM errors, and semantic/logic errors. The dataset contains approximately 22,000 training and 300 test samples, each annotated with buggy code, error types, error locations, corrected reference, and testbench. It supports three evaluation tasks---error classification, error localization, and error repair---and we report the repair success rate (Fix), measured as the fraction of buggy programs that the model successfully corrects as verified by testbench.

\subsubsection{GPU Optimization}

\paragraph{KernelBench.}
KernelBench~\cite{ouyang2025kernelbenchllmswriteefficient} evaluates LLMs on 250 PyTorch ML workloads across three levels: single operations at Level~1, operator sequences amenable to fusion at Level~2, and end-to-end architectures at Level~3. The model receives a PyTorch reference implementation and must produce an optimized version using any available tool. The benchmark introduces the $\text{fast}_p$ metric, measuring the fraction of tasks where the generated kernel is both correct and faster than threshold~$p$ over the PyTorch baseline. We report $\text{fast}_1$ separately for each level (L1, L2, L3).

\paragraph{TritonBench.}
TritonBench~\cite{li2025tritonbenchbenchmarkinglargelanguage} is the first comprehensive benchmark for Triton operator generation, targeting the Python-like GPU programming language adopted in vLLM and Liger-kernel. It features two evaluation tracks: TritonBench-G with 184 curated real-world operators from GitHub across five difficulty levels, and TritonBench-T with tasks aligned to PyTorch interfaces. For each track, we report two metrics: call accuracy (whether the generated code executes without error) and execution accuracy (whether the output further matches the reference implementation), yielding four metrics in total: G-call, G-exe, T-call, and T-exe.

\subsubsection{Code Optimization}

\paragraph{EmbedCGen.}
EmbedCGen is our proposed benchmark for bare-metal embedded C code generation targeting resource-constrained microcontrollers. It comprises 500 problems across five difficulty levels, ranging from basic peripheral control and register-level operations at lower levels to complex multi-peripheral integration, DMA mechanisms, and state machine coordination at higher levels. Generated code must satisfy HAL conventions and hard real-time constraints. Evaluation follows a strict serial pipeline, including code generation, cross-compilation via the ARM~GCC toolchain, and functional verification through the Renode system-level simulator. We report the average pass rate (Main) across all 500 problems.

\paragraph{SuperCoder.}
SuperCoder~\cite{wei2026supercoderassemblyprogramsuperoptimization} frames assembly superoptimization as an LLM task: given a C program and its \texttt{gcc -O3} output, the model must generate a semantically equivalent but faster assembly program. The benchmark contains 8,072 x86-64 assembly programs averaging 130 lines with loops, accompanied by test suites achieving 96.2\% line and 87.3\% branch coverage. Evaluation employs two metrics: accuracy assesses functional correctness through test suite validation, and average speedup quantifies performance gain relative to the compiler-optimized baseline.

\subsubsection{3D Modeling}

\paragraph{CAD-Coder.}
CAD-Coder~\cite{guan2025cadcodertexttocadgenerationchainofthought} reformulates text-to-CAD generation as producing executable CadQuery scripts, a Python-based parametric CAD language built on the OpenCascade kernel. The benchmark is constructed by transforming the Text2CAD dataset~\cite{khan2024text2cad} into 110K verified text-CadQuery-3D model triplets, stratified by quality into 8K high-quality, 70K medium-quality, and 32K hard cases, with an additional 1.5K chain-of-thought annotated samples. Evaluation employs two key metrics: compilation success rate measures the percentage of generated scripts that can be successfully executed to produce valid 3D geometry, while IoU quantifies the geometric fidelity by calculating the volumetric overlap between the generated voxels and the ground truth.

\definecolor{tablegray}{gray}{0.92}
\begin{table*}[!t]
    \centering
    \caption{Performance on chip design benchmarks. \modelname{} results are highlighted in gray.}
    \label{tab:chip_design_benchmarks}
    \resizebox{1.0\textwidth}{!}{
    \begin{tabular}{lr|c|c|cccccc|cc}
    \toprule
        \multirow{3}{*}{\textbf{Model}}
        & \multirow{3}{*}{\textbf{Size}}
        & \textbf{VeriScope}
        & \textbf{VeriRepair}
        & \multicolumn{6}{c|}{\textbf{RealBench}}
        & \multicolumn{2}{c}{\textbf{ArchXBench}} \\
        \cmidrule(lr){3-3} \cmidrule(lr){4-4} \cmidrule(lr){5-10} \cmidrule(lr){11-12}
        ~
        & ~
        & \multirow{2}{*}{\textbf{Score}}
        & \multirow{2}{*}{\textbf{Fix (\%)}}
        & \multicolumn{2}{c}{\textbf{System}}
        & \multicolumn{4}{c|}{\textbf{Module}}
        & \multirow{2}{*}{\textbf{$n$}}
        & \multirow{2}{*}{\textbf{$t$}} \\
        \cmidrule(lr){5-6} \cmidrule(lr){7-10}
        ~
        & ~
        & ~
        & ~
        & \textbf{Syn@1}
        & \textbf{Syn@5}
        & \textbf{Syn@1}
        & \textbf{Syn@5}
        & \textbf{Func@1}
        & \textbf{Func@5}
        & ~
        & ~ \\
        \midrule
        \multicolumn{12}{c}{\textbf{6B+ Models}} \\
        \midrule
        Qwen3.5-9B & 9B & 32.0 & 0.0 & 0.0 & 0.0 & 6.3 & 15.8 & 4.3 & 8.5 & 1.9 & 44.3 \\
        GPT-OSS-20B & 3.6/21B & \textbf{73.9} & \textbf{86.7} & 3.8 & 17.4 & \textbf{22.9} & \textbf{47.9} & 9.8 & \textbf{21.0} & \textbf{3.1} & \textbf{53.5} \\
        Qwen3.5-27B & 27B & 55.7 & 60.0 & \textbf{6.2} & \textbf{20.1} & 17.1 & 33.8 & \textbf{10.6} & 17.8 & 2.6 & 48.3 \\
        \midrule
        \multicolumn{12}{c}{\textbf{30B+ Models}} \\
        \midrule
        Qwen3-Coder-30B-A3B-Instruct & 3.3/30.5B & 66.2 & 76.7 & 0.0 & 0.0 & 23.0 & 35.2 & 5.2 & 8.2 & 2.4 & 37.3 \\
        Seed-OSS-36B-Instruct & 36B & 67.2 & 66.7 & 5.0 & 21.3 & 14.3 & 23.0 & 11.5 & 20.3 & 2.5 & 43.7 \\
        GPT-OSS-120B & 5.1/117B & 82.2 & 76.7 & 5.0 & 21.3 & 37.8 & 64.3 & 17.5 & 30.8 & 3.4 & \textbf{54.8} \\
        MiniMax-M2.5 & 10/230B & 75.1 & 66.7 & \textbf{23.8} & 48.5 & 17.2 & 38.4 & 6.9 & 15.9 & 2.9 & 46.0 \\
        GLM-4.7 & 32/355B & 81.2 & 63.3 & 12.5 & 24.6 & 25.4 & 46.9 & 11.6 & 21.2 & 3.2 & 51.4 \\
        GLM-5 & 40/744B & \textbf{83.2} & \textbf{90.0} & 2.5 & 11.2 & 22.2 & 43.4 & 12.2 & 22.6 & 3.1 & 53.2 \\
        Kimi-K2.5 & 32B/1T & 73.1 & 83.3 & 5.0 & 17.9 & 43.7 & 52.2 & 23.1 & 25.7 & \textbf{3.8} & 49.7 \\
        Kimi-K2-Instruct & 32B/1T & 82.4 & 76.7 & 6.2 & 26.2 & 50.1 & 70.1 & 22.2 & 28.3 & 2.9 & 44.9 \\
        Kimi-K2-Thinking & 32B/1T & 73.1 & 80.0 & 0.0 & 0.0 & 27.8 & 59.4 & 14.1 & 28.9 & 1.5 & 30.1 \\
        DeepSeek-V3.2 & 37/671B & 76.1 & 77.0 & 18.8 & \textbf{55.1} & 39.3 & 52.7 & 17.2 & 21.4 & 3.6 & 53.9 \\
        Qwen3.5-397B-A17B & 17/397B & 44.8 & 86.7 & 11.2 & 38.1 & 35.2 & 59.5 & 16.4 & 28.3 & 3.1 & 53.5 \\
        Qwen3-Coder-480B-A35B-Instruct & 35/480B & 80.8 & 76.7 & 0.0 & 0.0 & 28.9 & 39.5 & 14.8 & 20.6 & 3.0 & 43.9 \\
        \rowcolor{tablegray} \textbf{\modelname{}} & 32B & 80.7 & 80.0 & 10.0 & 23.7 & \textbf{74.8} & \textbf{83.3} & \textbf{62.7} & \textbf{70.5} & 3.4 & 51.0 \\
        \midrule
        \multicolumn{12}{c}{\textbf{Closed-APIs Models}} \\
        \midrule
        Claude-Sonnet-4.6 & \faLock{} & 87.7 & 83.3 & 41.2 & 50.0 & 69.2 & 77.7 & 33.5 & 37.2 & 4.4 & 58.2 \\
    \bottomrule
    \end{tabular}
    }
\end{table*}

\begin{table*}[!t]
    \centering
    \caption{Performance on GPU optimization, code optimization, and 3D modeling benchmarks. \modelname{} results are highlighted in gray.}
    \label{tab:other_industrial_benchmarks}
    \resizebox{1.0\textwidth}{!}{
    \begin{tabular}{lr|cc|c|cc|cccc|ccc}
    \toprule
        \multirow{2}{*}{\textbf{Model}}
        & \multirow{2}{*}{\textbf{Size}}
        & \multicolumn{2}{c|}{\textbf{CAD-Coder}}
        & \textbf{EmbedCGen}
        & \multicolumn{2}{c|}{\textbf{SuperCoder}}
        & \multicolumn{4}{c|}{\textbf{TritonBench}}
        & \multicolumn{3}{c}{\textbf{KernelBench}} \\
        \cmidrule(lr){3-4} \cmidrule(lr){5-5} \cmidrule(lr){6-7} \cmidrule(lr){8-11} \cmidrule(lr){12-14}
        ~
        & ~
        & \textbf{Comp.}
        & \textbf{IoU}
        & \textbf{Main (\%)}
        & \textbf{Acc. (\%)}
        & \textbf{Spd.}
        & \textbf{G-call (\%)}
        & \textbf{G-exe (\%)}
        & \textbf{T-call (\%)}
        & \textbf{T-exe (\%)}
        & \textbf{L1}
        & \textbf{L2}
        & \textbf{L3} \\
        \midrule
        \multicolumn{14}{c}{\textbf{6B+ Models}} \\
        \midrule
        Qwen3.5-9B & 9B & 2.0 & 0.0 & 10.0 & \textbf{36.0} & 1.0$\times$ & 2.7 & \textbf{100.0} & 3.6 & \textbf{100.0} & 0.0 & 0.0 & 2.0 \\
        GPT-OSS-20B & 3.6/21B & 2.0 & \textbf{2.0} & \textbf{30.6} & 16.0 & 1.0$\times$ & 2.2 & \textbf{100.0} & 1.2 & \textbf{100.0} & \textbf{5.1} & \textbf{10.0} & 2.0 \\
        Qwen3.5-27B & 27B & \textbf{4.0} & 0.3 & 9.6 & 6.0 & \textbf{1.9}$\times$ & \textbf{5.4} & \textbf{100.0} & \textbf{25.9} & 97.7 & \textbf{5.1} & 6.0 & \textbf{4.0} \\
        \midrule
        \multicolumn{14}{c}{\textbf{30B+ Models}} \\
        \midrule
        Qwen3-Coder-30B-A3B-Instruct & 3.3/30.5B & 0.0 & 0.0 & 15.4 & 50.0 & 1.0$\times$ & 8.7 & \textbf{100.0} & 24.1 & 67.5 & 2.0 & 0.0 & 0.0 \\
        Seed-OSS-36B-Instruct & 36B & 2.0 & 2.0 & 10.2 & 8.0 & 1.1$\times$ & 1.6 & \textbf{100.0} & 1.8 & \textbf{100.0} & 1.0 & 2.0 & 2.0 \\
        GPT-OSS-120B & 5.1/117B & 4.0 & 1.9 & 17.8 & 8.0 & 1.2$\times$ & 3.8 & 85.7 & 12.7 & 95.2 & 6.1 & 15.0 & 2.0 \\
        MiniMax-M2.5 & 10/230B & 4.0 & 0.4 & 22.2 & 20.0 & 3.5$\times$ & 5.4 & \textbf{100.0} & 15.1 & \textbf{100.0} & 7.1 & 14.0 & 8.0 \\
        GLM-4.7 & 32/355B & 12.0 & 6.0 & 89.6 & 20.0 & 8.6$\times$ & 3.3 & \textbf{100.0} & 6.0 & \textbf{100.0} & 8.1 & 19.0 & 0.0 \\
        GLM-5 & 40/744B & 38.0 & 18.8 & \textbf{90.2} & 54.0 & 1.87$\times$ & 1.6 & \textbf{100.0} & 1.2 & \textbf{100.0} & 16.2 & 23.0 & 4.0 \\
        Kimi-K2.5 & 32B/1T & 40.0 & 12.1 & 81.0 & 46.0 & 1.9$\times$ & 12.5 & \textbf{100.0} & 7.8 & \textbf{100.0} & 13.1 & 23.0 & 6.0 \\
        Kimi-K2-Instruct & 32B/1T & 2.0 & 1.1 & 69.6 & 12.0 & 1.1$\times$ & 15.8 & 96.5 & 13.9 & 91.3 & 6.1 & 0.0 & 0.0 \\
        Kimi-K2-Thinking & 32B/1T & 48.0 & 20.0 & - & 24.0 & 1.2$\times$ & 17.4 & \textbf{100.0} & 19.9 & 84.8 & 9.1 & 16.0 & 4.0 \\
        DeepSeek-V3.2 & 37/671B & 14.0 & 4.6 & 84.4 & 30.0 & 1.8$\times$ & 19.6 & \textbf{100.0} & 18.1 & 13.3 & 3.0 & 0.0 & 0.0 \\
        Qwen3.5-397B-A17B & 17/397B & 36.0 & 14.2 & 17.8 & 34.0 & 1.2$\times$ & 7.6 & \textbf{100.0} & 16.3 & 92.6 & 4.0 & 10.0 & 0.0 \\
        Qwen3-Coder-480B-A35B-Instruct & 35/480B & 10.0 & 4.7 & 9.0 & 64.0 & 2.0$\times$ & \textbf{20.1} & \textbf{100.0} & \textbf{31.9} & 56.6 & 3.0 & 6.0 & 0.0 \\
        \rowcolor{tablegray} \textbf{\modelname{}} & 32B & \textbf{82.0} & \textbf{53.5} & 35.2 & \textbf{91.0} & 1.3$\times$ & 18.5 & \textbf{100.0} & 19.3 & 93.8 & \textbf{22.2} & \textbf{36.0} & \textbf{14.0} \\
        \midrule
        \multicolumn{14}{c}{\textbf{Closed-APIs Models}} \\
        \midrule
        Claude-Sonnet-4.6 & \faLock{} & 77.0 & 32.4 & 79.0 & 88.0 & 4.6$\times$ & 28.8 & 98.1 & 41.6 & 1.4 & 11.1 & 28.0 & 2.0 \\
    \bottomrule
    \end{tabular}
    }
\end{table*}

\subsection{Main Results}

\paragraph{Results on General Code Benchmarks}
As shown in \autoref{tab:code_generation_1}, \autoref{tab:code_reasoning_efficiency_sql} and \autoref{tab:agentic_combined}, \modelname{} as a 32B dense model achieves broadly competitive results with significantly larger open-weight models across code generation, code reasoning, and Text2SQL, while ranking first among all open-weight baselines on SWE-bench Verified, Mind2Web, and $\tau^2$-bench. Although a gap remains on certain reasoning and efficiency benchmarks compared to the largest MoE models, \modelname{} delivers strong overall performance given its compact size.

\paragraph{Results on Industrial Code Benchmarks}
As shown in \autoref{tab:chip_design_benchmarks} and \autoref{tab:other_industrial_benchmarks}, \modelname{} demonstrates clear advantages on industrial code tasks. On chip design, it achieves the best open-weight results on RealBench module-level tasks by a wide margin, while remaining competitive on VeriScope and VeriRepair. On 3D modeling and GPU optimization, \modelname{} leads all open-weight baselines on CAD-Coder and KernelBench across all three levels, and even surpasses the proprietary Claude-Sonnet-4.6 on CAD-Coder IoU and KernelBench L1/L2/L3. While tasks such as EmbedCGen and SuperCoder remain challenging, the results overall confirm that \modelname{} provides strong industrial code capabilities.
benchmarksabout  that go beyond standard functional testinge at L1 through sequential circuitsand -based testingcode that fails to compile receives a score of 0, code that compiles successfully but fails tests receives 50 points, and code that passes all tests receives the full 100 points.provides, covering domainsThe benchmarkuses  evaluation, reporting results as $(n, t)$ pairs where $n$ is the number of compilable candidates out of five attempts, and $t$ is the testbench pass rate of the best candidate.covering acrossEemploys
\section{Related Work}

\subsection{Large Language Models for Code}

The code intelligence landscape has evolved rapidly, building on a rich history of code-oriented LLM research~\cite{nijkamp2023codegen2,zheng2023codegeex,lozhkov2024starcoder2stackv2,roziere2024codellama,deepseek2024coder,hui2024qwen2,huang2025opencoder,seedcoder,li2025aixcoder,yicoder2024,stabilityai2023stablecode,agarwal2025gpt,seed2025seed-oss}.
On the open-weight side, the DeepSeek series~\cite{liu2025deepseekv32,deepseekv31terminus}, the GLM series~\cite{glm47,zeng2026glm}, the Kimi family~\cite{team2025kimi2,team2026kimi25}, the MiniMax series~\cite{minimax-m2,minimax-m21,minimax-m25}, and the Qwen ecosystem~\cite{qwen3.5,qwen3next,qwen3coder,qwen_qwen3_coder_next_tech_report} have demonstrated strong performance on both standard code generation benchmarks and agentic coding tasks.
On the closed-source side, the Claude series~\cite{claude45,claude45opus,claude46,claude46opus}, the GPT series~\cite{openai2025gpt5developers,gpt5-3-codex,gpt5-4}, and the Gemini series~\cite{gemini3pro,gemini31pro} continue to push the frontier, excelling at complex code reasoning, multi-step planning, and tool-augmented agentic workflows.
A rich set of benchmarks has been developed in parallel to track these advances, covering function-level correctness~\cite{chen2021codex,evalplus}, multilingual evaluation~\cite{MultiPL-E,ziyao2023xcodeeval}, complex real-world tasks~\cite{zhuo2024bigcodebench,liu2024fullstackbench,du2025swe}, contamination-free assessment~\cite{jain2024livecodebenchholisticcontaminationfree}, human preference alignment~\cite{codearena}, and agentic coding capabilities~\cite{jimenez2024swebench,tbench2025,deng2023mind2web,patil2025bfcl,yao2024tau}.

Despite this rapid progress, existing LLMs for code are predominantly trained and evaluated on general software engineering tasks such as algorithm implementation, web development, and scripting.
The specific demands of industrial programming domains, including hardware description languages, high-performance GPU kernel programming, compiler optimization, and computer-aided design, remain largely underexplored.

\subsection{Industrial Code Intelligence}

Industrial software development involves specialized domains where code LLMs must handle domain-specific language syntax, hardware-aware optimization, and strict functional correctness, posing unique challenges beyond general-purpose programming. Recent work has begun to address these challenges, though mostly in a domain-specific manner.

Chip design has received the most attention. Early efforts such as VeriGen~\cite{thakur2023verigen} and RTLCoder~\cite{liu2024rtlcoder} fine-tuned LLMs for RTL code generation, with subsequent work expanding to Verilog debugging, multi-modal synthesis, and verification~\cite{ea2023benchmarking,wang2025veridebug,chang2024natural,yubeaton2025verithoughts,yang2025large}. Recent methods increasingly adopt reinforcement learning with hardware-specific rewards, including CodeV-R1~\cite{zhu2025qimeng}, VeriReason~\cite{wang2025verireason}, and ReasoningV~\cite{qin2025reasoningv}. Corresponding benchmarks such as VerilogEval~\cite{liu2023verilogeval,pinckney2024revisitingverilogevalnewerllms}, RealBench~\cite{jin2025realbench}, ArchXBench~\cite{purini2025archxbench}, CVDP~\cite{pinckney2025comprehensive}, MetRex~\cite{abdelatty2025metrex}, and VeriBench~\cite{agarwal2025veribench} provide increasingly rigorous evaluation for hardware design. For embedded systems, EmbedGenius~\cite{yang2024embedgenius} and EmbedAgent~\cite{xu2025embedagent} explore automated IoT development and benchmarking~\cite{englhardt2024exploring,quan2025sensorbench,delorenzo2024creativeval,babiuch2026benchmarking}.

GPU kernel optimization has also seen rapid progress~\cite{fischeskernelllm,mikasa2026improving,yu2026towards}. Kevin~\cite{baronio2025kevin}, ConCuR~\cite{kong2025concur}, and CUDA Agent~\cite{dai2026cuda} apply reinforcement learning to CUDA kernel generation, while AscendKernelGen~\cite{cao2026ascendkernelgen} and Dr.Kernel~\cite{liu2026dr} further extend to NPU kernels and profiling-based reward design, with KernelBench~\cite{kernelbench} and TritonBench~\cite{tritonbench} serving as standard benchmarks. For compiler optimization, SuperCoder~\cite{wei2025supercoder}, LLM-Vectorizer~\cite{taneja2025llm}, LLM Compiler~\cite{cummins2025llm}, and LLM-VeriOpt~\cite{fang2026llm} target tasks from assembly superoptimization to LLVM-IR peephole optimization. In 3D modeling, CAD-Coder~\cite{guan2025cad} and ReCAD~\cite{li2025recad} generate parametric CAD scripts from text and images~\cite{niu2025intent,zhou2025cad}, while STEP-LLM~\cite{shi2026step} and BrepCoder~\cite{kim2026brepcoder} target industry-standard STEP and B-rep formats.

While these efforts have significantly advanced individual domains, each model or benchmark addresses a single industrial sub-domain, leading to fragmented coverage. In contrast, \modelname{} is a unified industrial code LLM that spans multiple industrial programming domains, bridging the gap between general-purpose code LLMs and industrial software development.

\section{Analysis}
\subsection{Error Analysis}
\label{sec:error_analysis}

\begin{figure*}[t!]
    \centering
    \includegraphics[width=\textwidth]{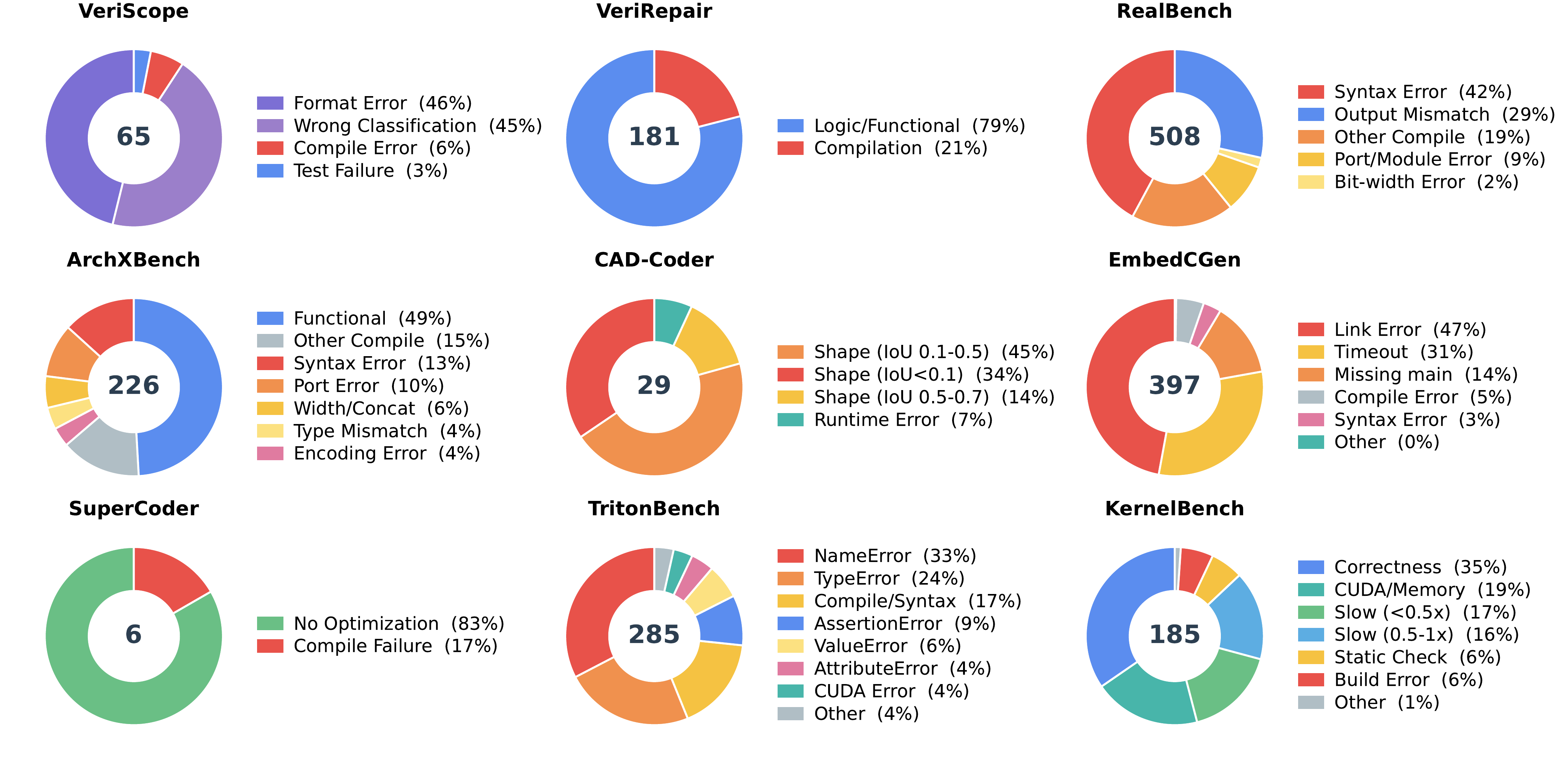}
    \caption{Error distribution of \modelname{} across 9 industrial benchmarks. The center number indicates total failures per benchmark. Errors are color-coded by type: reds/oranges for compilation and syntax errors, blues for functional/logic errors, purples for format errors, greens for performance issues.}
    \label{fig:error_analysis}
\end{figure*}

Despite strong overall performance, \modelname{} still exhibits systematic failure patterns that reveal the remaining challenges of industrial code generation. We manually inspect all 1,882 failure cases across the 9 industrial benchmarks and categorize them into five recurring error themes. \autoref{fig:error_analysis} presents the error distribution for each benchmark.

The most prevalent failure category is compilation and syntax errors, which dominate Verilog generation tasks: 71\% of RealBench failures involve malformed literals, incorrect port declarations, or bit width mismatches, and ArchXBench shows a similar pattern with 51\% syntax errors including misuse of named ports on primitive gates and signed literals of indefinite width. Beyond syntax, knowledge of industrial APIs remains incomplete---on EmbedCGen, 47\% of failures are linker errors from referencing undefined or incorrectly typed HAL/CMSIS functions, while TritonBench sees 33\% \texttt{NameError}s and 24\% \texttt{TypeError}s from incorrect Triton API usage. A further 46\% of VeriScope failures are unparseable outputs where the model ignores the required structured format entirely. Together, these errors suggest that \modelname{} has acquired broad domain vocabulary but has not fully internalized the precise syntactic rules and library constraints of industrial languages.

When code compiles, functional correctness becomes the dominant bottleneck. VeriRepair illustrates this most sharply: 79\% of its failures produce code that compiles but fails test cases, pointing to subtle logic errors in repair reasoning rather than mistakes at the surface level. ArchXBench echoes this with 49\% functional failures where generated Verilog simulates incorrectly, and CAD-Coder failures are almost entirely geometric (93\%)---most stemming from a systematic misinterpretation of Euler angle conventions that rotates extrusions around the wrong axis. These cases share a common structure: the model produces superficially plausible code that breaks under precise numerical or state machine semantics, revealing the difficulty of reasoning about correctness in domains where small mistakes propagate into observable failures.

Benchmarks that measure optimization expose a third gap: generating correct code is necessary but not sufficient for industrial tasks. On KernelBench, 33\% of failures produce functionally correct but insufficiently fast GPU kernels, and on SuperCoder 83\% of failures reduce to the model simply copying the input x86-64 assembly without modification. Collectively, these failure patterns indicate that industrial code intelligence demands more than general code training can provide: closing the gaps will require data curation targeting rare APIs and hardware semantics, training with verification signals in the loop, and explicit reasoning about low-level performance. \textbf{There is still a long journey to apply LLMs in realistic industrial scenarios.}

\begin{figure*}[t!]
    \centering
    \includegraphics[width=\textwidth]{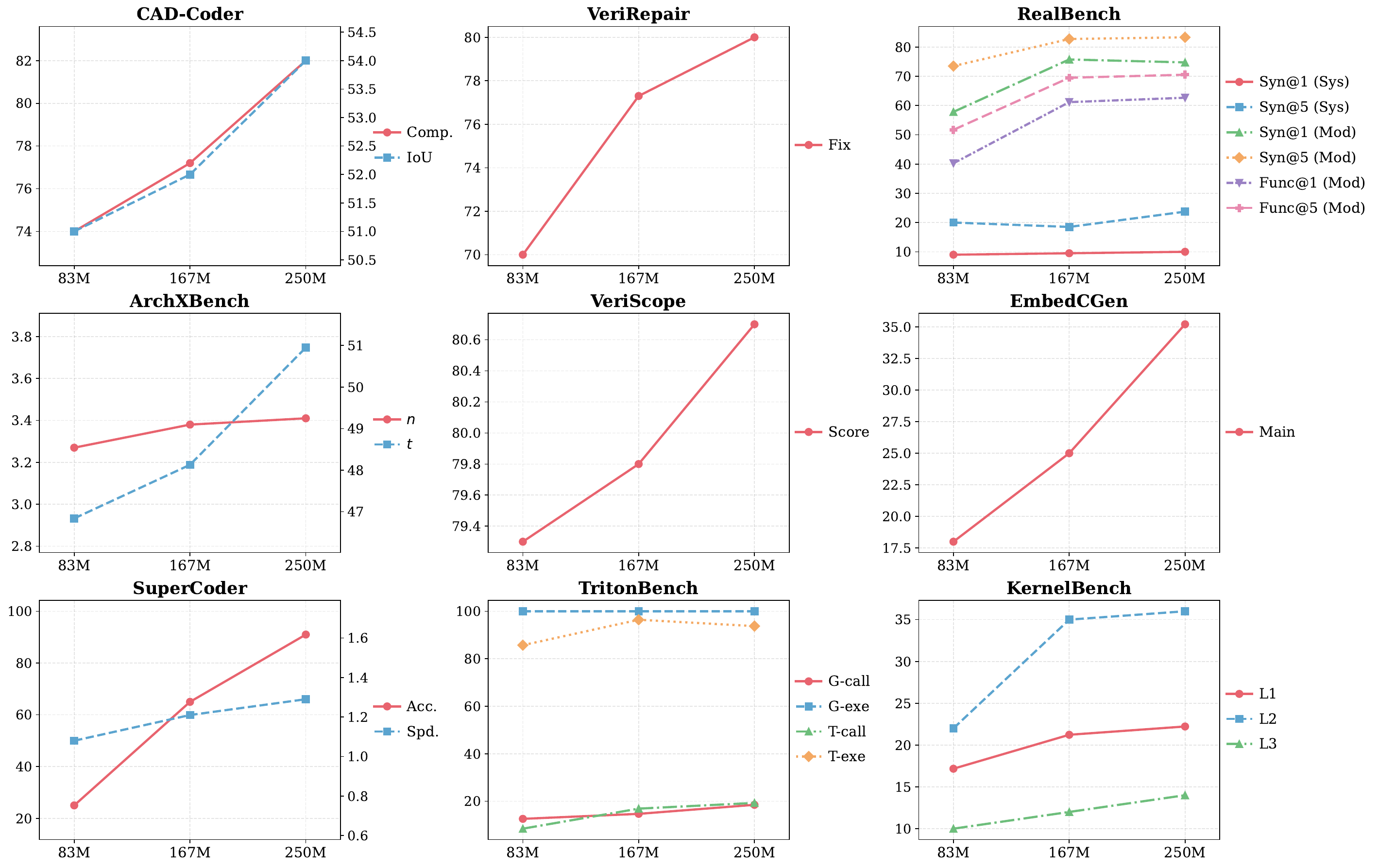}
    \caption{Performance across nine industrial code benchmarks as SFT data scales from 83M to 250M tokens.}
    \label{fig:efftct_of_stage}
\end{figure*}

\subsection{Effects of different Stages}

To understand how the volume of industrial SFT data affects downstream performance, we train three checkpoints on 83M, 167M, and 250M tokens respectively and evaluate each across all nine benchmarks. \autoref{fig:efftct_of_stage} summarizes these trends.
The majority of benchmarks show consistent improvements as the data scale grows, confirming that core industrial coding and architectural reasoning capabilities benefit from larger-scale fine-tuning.
Only a few individual sub-metrics on RealBench and TritonBench exhibit minor regressions at the 250M stage, yet they still remain above or close to the 83M baseline, suggesting that verification-related understanding may saturate early with a smaller high-quality SFT dataset. Overall, these findings confirm that scaling industrial SFT data is a reliable driver for performance, with only marginal fluctuations observed at the largest scale.

\section{Conclusion}
In this paper, we have presented \modelname{}, a code foundation model that bridges the gap between general code intelligence and the stringent demands of industrial software development. Through a systematic three-stage Code-Flow training pipeline, \modelname{} effectively acquires reasoning capabilities about hardware behavior and industrial constraints without sacrificing general programming performance, while offering both instruction tuning and analytical reasoning within a unified framework. Extensive evaluations across 14 general benchmarks and 9 industrial benchmarks covering chip design, embedded systems, GPU optimization, and compiler optimization demonstrate that \modelname{} achieves competitive performance with leading models on general tasks while establishing strong baselines in industrial domains. Our ablation studies further reveal that repository transition data outperforms static snapshots for planning signals, mid-training trajectory data serves as a critical scaffold for robustness under distribution shift, and training with multi turn trajectories conditioned on execution feedback unlocks capabilities absent in standard SFT.

\clearpage
\newpage

\bibliography{ref}

\newpage
\appendix

\tcbset{
  appendixbox/.style={
    breakable,
    enhanced,
    pad at break*=2mm,
    extras first={
      overlay={\draw[tcbcolframe, line width=0.5mm]
        (frame.south west) -- (frame.south east);}
    },
    extras middle={
      overlay={
        \draw[tcbcolframe, line width=0.5mm]
          (frame.north west) -- (frame.north east);
        \draw[tcbcolframe, line width=0.5mm]
          (frame.south west) -- (frame.south east);
      }
    },
    extras last={
      overlay={\draw[tcbcolframe, line width=0.5mm]
        (frame.north west) -- (frame.north east);}
    },
  }
}

\label{sec:appendix}

\section{Pre-training Details}
\label{sec:pretraining_appendix}

\subsection{Model Architecture}

\modelname{} adopts a standard decoder-only Transformer architecture. The detailed model configurations are provided in \autoref{tab:model-config}.

\subsection{Pre-training Data}
\label{subsec:pretrain-data}

Existing large-scale code corpora, such as The Stack v2~\cite{lozhkov2024starcoder2stackv2}, StarCoderData~\cite{li2023starcoder}, and The Pile~\cite{gao2020pile}, are predominantly composed of web-oriented languages (JavaScript, TypeScript, Python for web frameworks), while industrial code remains underrepresented. Verilog and VHDL occupy only a negligible fraction of total tokens. CUDA and Triton kernels are not separately categorized and are often mixed into general C/C++ code. Low-level systems code such as device drivers and firmware is difficult to distinguish from ordinary C programs. This imbalance directly limits the industrial capabilities of models trained on such data. To address this, we construct a data pipeline for collecting, cleaning, and refining industrial code.

\subsubsection{Data Collection.}
We collect industrial code from three complementary sources.

\paragraph{Repository-level recall.}
We define a domain taxonomy covering digital circuit design, high-performance computing, embedded systems, and CAD automation, and retrieve relevant code from large-scale public repositories through a three-step recall strategy with increasing coverage. In the first step, we apply rule-based filtering using file extensions, directory naming conventions, and domain-specific keywords (e.g., \texttt{endmodule}/\texttt{posedge} for Verilog, \texttt{\_\_global\_\_}/\texttt{<<<...>>>} for CUDA, \texttt{\#pragma HLS} for FPGA synthesis) to collect files with obvious industrial characteristics. In the second step, we train a FastText classifier on a manually labeled seed set to recall additional industrial code that shares similar statistical patterns but does not match simple rules. In the third step, we employ a domain-adapted semantic encoder to retrieve samples that lack both lexical and statistical features captured by the previous two steps, such as hardware-related C/C++ libraries and low-level driver code that are syntactically indistinguishable from general-purpose programs.

\paragraph{Technical literature via OCR.}
We apply OCR to extract code snippets and structured content from technical books and hardware reference manuals, recovering domain knowledge that is largely absent from public code repositories.

\paragraph{Domain-specific web sources.}
We further collect data from technical forums, vendor documentation, and engineering reports, which provide practical code samples and usage patterns not covered by the above two sources.

\subsubsection{Data Cleaning.}
The collected data undergoes multi-step cleaning. We first filter out files with restrictive licenses, remove personally identifiable information and embedded credentials, and discard trivially invalid files such as empty stubs, auto-generated boilerplate, and binary artifacts. Deduplication is then performed at four levels: exact hash matching, token-level near-duplicate detection via MinHash LSH~\cite{lozhkov2024starcoder2stackv2}, repository-level fork consolidation, and cross-source deduplication. We also apply lightweight domain-specific checks, including syntax parsing for Verilog and SystemVerilog, header consistency checks for C/C++, and format validation for GPU kernel configurations. Samples that fail these checks are routed to the refinement step for repair or removed.

\subsubsection{Data Refinement.}
We refine the cleaned data in two steps. The first step normalizes surface-level formatting by unifying style conventions, standardizing module boundaries, and removing noise such as commented-out dead code and outdated compatibility blocks. The second step adds structured annotations, including cross-file dependency resolution, platform and constraint metadata, and natural-language descriptions at the function, module, and project levels for code-text alignment. All refined samples are verified through AST comparison and re-compilation to ensure correctness.

\subsection{Training}
\label{subsec:pretrain-training}

We train \modelname{} on 4,096 GPUs with autoregressive language modeling and fill-in-the-middle (FIM) completion~\cite{gong2025structureawarefillinthemiddlepretrainingcode, guo2024deepseek}, using a constant learning rate of $3\times10^{-4}$ and a global batch size of 2,048 for a total of 15T tokens. Training data follows a curriculum schedule~\cite{bengio2009curriculum} that progresses from function-level, single-file samples to multi-file, project-level data. A proportion of general-purpose code and text is retained in every batch to preserve general programming capabilities. We periodically evaluate on held-out industrial benchmarks and adjust data sampling weights accordingly.


\section{Mid-Training Loss Curves}
\label{app:loss-curves}

\begin{figure}[h!]
    \centering
    \begin{subfigure}[b]{0.48\textwidth}
        \centering
        \includegraphics[width=\textwidth]{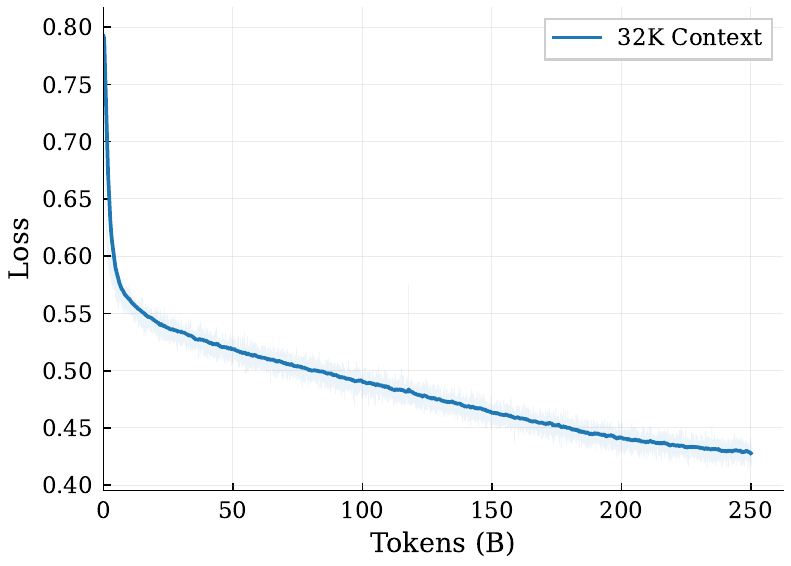}
        \caption{Stage 2.1: 8K to 32K context extension}
        \label{fig:loss-32k}
    \end{subfigure}
    \hfill
    \begin{subfigure}[b]{0.48\textwidth}
        \centering
        \includegraphics[width=\textwidth]{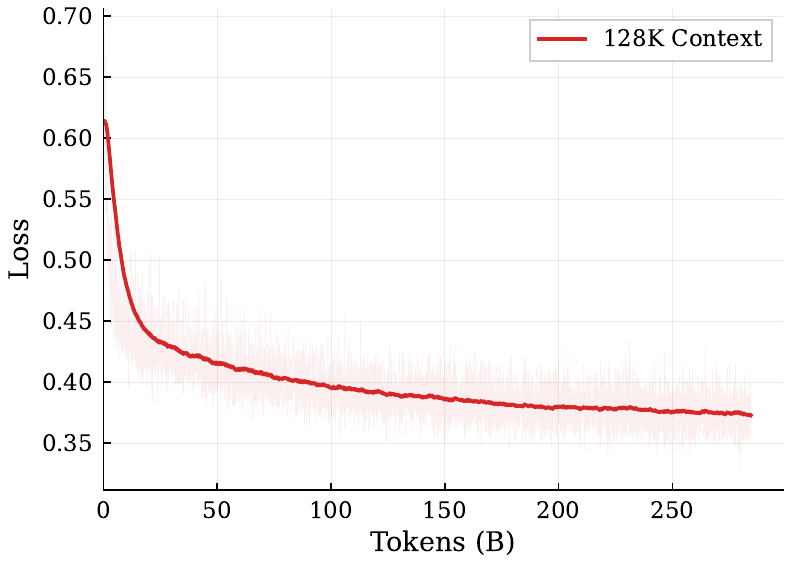}
        \caption{Stage 2.2: 32K to 128K context extension}
        \label{fig:loss-128k}
    \end{subfigure}
    \caption{Training loss curves for both stages of mid-training. (a) Stage 2.1 shows consistent convergence with cosine learning rate decay. (b) Stage 2.2 reflects the graduated warm-up strategy for long-sequence samples.}
    \label{fig:loss-curves}
\end{figure}

This appendix presents the training loss curves for the two stages of mid-training described in \autoref{sec:mid-training}. \autoref{fig:loss-curves} shows the training dynamics for both Stage~2.1 (8K to 32K extension) and Stage~2.2 (32K to 128K extension).
In Stage~2.1, the loss curve demonstrates stable convergence with the cosine learning rate decay schedule, indicating successful adaptation to longer context windows while maintaining training stability.
In Stage~2.2, the graduated warm-up strategy for long-sequence samples (starting at 10\% and increasing to 50\%) is reflected in the loss dynamics, with stable convergence achieved after the warm-up period.

\section{Mid-Training Details}
\label{sec:mid-training_appendix}

Industrial hardware and systems development demands capabilities beyond what general-purpose code models provide.
Production hardware designs: RTL circuits, GPU kernels, systems code, FPGA synthesis: involve domain-specific languages (Verilog, C/C++, Triton), stringent timing and resource constraints,
and verification methodologies rarely seen in general software repositories~\cite{Gao_2025,diggs2024leveragingllmslegacycode}.
Models trained on web application code lack the structured reasoning for hardware scenarios, often producing outputs that violate timing constraints or introduce synthesis errors.

Mid-training bridges this gap by continuing pretraining on hardware-aware industrial data with targeted objectives~\cite{mo2025midtraininglargelanguagemodels}, transforming the base model into an \emph{industrial-grade hardware-aware code foundation model} capable of reasoning about timing, resource optimisation, and performance-critical system implementations.

\subsection{Progressive Context Extension Strategy}
\label{subsec:training-strategy}

Our mid-training adopts a two-stage progressive curriculum extending the model's context from pretrained 8K tokens to 128K tokens.

\paragraph{Stage 2.1: 8K to 32K Extension.}
Stage~2.1 extends from the pretrained 8K checkpoint to 32K tokens, targeting file-level tasks: completing RTL modules, infilling kernel functions, and generating testbenches. We train directly on sequences up to 32K without positional encoding interpolation. The data mixture emphasises reasoning QA (40\%), agent trajectories (20\%), commits (15\%), industrial artefacts (15\%), and FIM (10\%), with cosine learning rate decay to avoid catastrophic forgetting. The training loss curve demonstrates stable convergence throughout this stage (see Figure~\ref{fig:loss-32k} in Appendix~\ref{app:loss-curves}).

\paragraph{Stage 2.2: 32K to 128K Extension.}
Stage~2.2 extends to 128K tokens, unlocking long-context capabilities: extended debugging sessions, large hardware projects with cross-module dependencies, and multi-file refactoring campaigns. Long-sequence samples (>32K) are introduced via graduated warm-up: starting at 10\% and increasing linearly to 50\% over the first quarter of training steps. Data replay (5--10\% from Stage~2.1) preserves file-scale performance. The learning rate is reset and follows a second cosine decay. Training dynamics reflect the graduated warm-up strategy with stable convergence after the warm-up period (see Figure~\ref{fig:loss-128k} in Appendix~\ref{app:loss-curves}).

\subsection{Industrial Data Synthesis and Curation}
\label{subsec:data}

Our training data consists of two complementary sources: curated industrial code data and synthetically generated industrial reasoning QA. While curated data provides authentic examples from real-world codebases, synthetic data enables systematic coverage of specialised industrial scenarios underrepresented in public repositories.

\subsubsection{Synthetic Industrial Code QA}
\label{subsubsec:synthetic-qa}

\paragraph{Motivation.}
Industrial hardware and systems engineering spans diverse domains: digital circuit design (RTL/Verilog), GPU computing (Triton operators, CUDA kernels), systems programming (C/C++/Rust kernels), FPGA synthesis (HLS), CAD tool integration, and embedded systems—each with domain-specific challenges, timing constraints, resource budgets, and verification methodologies. Publicly available code repositories exhibit strong distributional bias toward web applications and high-level software, leaving substantial gaps in hardware and low-level systems coverage. Synthetic data generation addresses this gap by systematically constructing reasoning QA pairs that reflect the full spectrum of hardware-aware industrial contexts, ensuring the model's capabilities generalise beyond the limited distribution of organic training data.

\paragraph{Three-Stage Synthesis Pipeline.}
Our synthesis pipeline operates in three stages designed to produce industrially grounded, factually correct reasoning data:

\textbf{(i) Industrial scenario specification.}
We identify concrete industrial engineering contexts through consultation with practising hardware and systems engineers and systematic analysis of industrial design verification reports.
Target scenarios include: debugging timing violations in digital circuits (setup/hold time failures); optimising Verilog RTL designs for area-power-performance trade-offs; implementing high-performance GPU operators in Triton with memory coalescing and shared memory tiling; verifying functional correctness of C/C++ kernels with formal methods or symbolic execution; resolving synthesis errors in HLS-generated FPGA designs; diagnosing memory safety issues (buffer overflows, use-after-free) in systems code; analysing waveform traces to localise hardware bugs; and implementing backward-compatible hardware-software interface changes. Each scenario is specified with concrete industrial characteristics: typical design scale, hardware description language, timing/resource constraints, verification requirements, and common
failure modes.

\textbf{(ii) Seed code generation.}
For each scenario, we generate or extract representative code fragments that exhibit target industrial
characteristics in hardware and systems domains.
These seed codes are constructed to reflect realistic hardware design patterns, timing-critical
implementations, memory management idioms, and domain-specific conventions.
We prioritise code patterns that appear frequently in industrial practice but are underrepresented
in academic benchmarks: finite state machines in RTL designs, pipelined datapath implementations,
GPU kernel memory access patterns (coalesced vs. strided), SIMD vectorisation in C kernels,
interrupt-driven embedded firmware, and CAD tool scripting for design automation.

\textbf{(iii) QA pair synthesis with automated verification.}
For each seed code, we synthesise question-answer pairs where questions probe industrial reasoning
competencies—root-cause analysis, performance diagnosis, correctness verification, refactoring
trade-offs, and impact analysis—and answers provide step-by-step reasoning traces following the
chain-of-thought paradigm~\cite{wei2023chainofthoughtpromptingelicitsreasoning}.
Crucially, all synthesised reasoning traces undergo \emph{automated verification} to guarantee
factual correctness:
\begin{itemize}[leftmargin=*,noitemsep,topsep=2pt]
  \item \textbf{Code execution validation}: computational steps are verified by executing code
    snippets and checking outputs against expected results;
  \item \textbf{Static analysis}: semantic properties (type correctness, control-flow reachability,
    data-flow dependencies) are confirmed via static analysis tools;
  \item \textbf{Logical consistency checks}: reasoning chains are validated for logical soundness,
    ensuring each step follows from previous steps without contradictions.
\end{itemize}
This verification layer distinguishes our approach from naive prompt-based synthesis.
It ensures the model trains only on industrially grounded, factually correct reasoning patterns.
This eliminates a critical failure mode: the model learning fluent but semantically incorrect
reasoning that appears correct superficially but causes failures in production systems.

\paragraph{Coverage and Scale.}
Our synthetic QA corpus systematically covers industrial scenarios across:
\textbf{domains}: RTL design (Verilog/SystemVerilog), GPU computing (Triton/CUDA), systems
programming (C/C++/Rust kernels), FPGA synthesis (HLS/Vivado), CAD automation (Tcl/Python scripting),
embedded firmware;
\textbf{reasoning types}: timing analysis (setup/hold), resource optimisation (area/power/latency),
memory safety verification, concurrency correctness, numerical precision analysis;
and \textbf{operational contexts}: RTL verification (waveform debugging, assertion failures),
kernel performance tuning, synthesis error resolution, hardware-software co-design, and
design-for-test integration.
The resulting synthetic data complements curated samples by filling coverage gaps and providing
dense training signal for the structured reasoning patterns that hardware-aware industrial
engineering demands.

\subsubsection{Curated Industrial Code Data}
\label{subsubsec:curated-data}

To complement synthetic QA, we curate authentic industrial code data spanning four categories:

\paragraph{Agent Trajectories.}
Multi-step debugging and repair trajectories following Thought-Action-Observation
cycles~\cite{yang2024sweagentagentcomputerinterfacesenable}, capturing closed-loop reasoning with
tool feedback from hardware simulators, synthesis tools, C/C++ compilers, and formal verification
engines.
These trajectories teach the model to diagnose errors from industrial signals (synthesis warnings,
simulation mismatches, timing violations, memory sanitiser reports) and iterate toward correct
solutions.

\paragraph{Code Commits.}
Version-control commits pairing developer intent (commit message) with before-after code states,
curated from large-scale open-source repositories~\cite{Liu_2024}.
To enhance coverage of hardware and systems scenarios, we supplement general software commits with
hardware-characteristic samples: RTL optimisation for timing closure, memory hierarchy refactoring
in kernels, ABI compatibility preservation, and hardware bug fixes with testbench updates.

\paragraph{Industrial Code Artefacts.}
Beyond standard source code, we include auxiliary artefacts that reflect the operational context
of professional development: test suites, API specifications, configuration files, and log
samples~\cite{srivatsa2024surveyusinglargelanguage,alshahwan2024automatedunittestimprovement}.
For hardware and systems domains specifically, we incorporate hardware testbenches (SystemVerilog/UVM),
timing constraints (SDC), synthesis scripts, GPU profiling traces, and memory sanitiser logs.

\paragraph{File-Level Fill-in-the-Middle (FIM).}
Structure-aware FIM samples targeting function bodies, class methods, and code
blocks~\cite{guo2024deepseekcoderlargelanguagemodel,gong2025structureawarefillinthemiddlepretrainingcode}.
AST-guided masking respects structural boundaries to ensure syntactic validity.
For hardware and systems code, we apply domain-specific masking strategies: respecting always blocks
and port declarations in Verilog, and targeting semantically coherent units (loop bodies, SIMD
intrinsic blocks) in C/C++/Triton kernels.

\paragraph{Data Mixture.}
Stage~2.1 weights synthetic reasoning QA most heavily (40\%), complemented by agent trajectories
(20\%), commits (15\%), industrial artefacts (15\%), and FIM (10\%).
Stage~2.2 shifts weight toward long-context data: agent trajectories increase to 30\%, FIM to 25\%,
while maintaining reasoning QA at 25\% and other categories at 10\% each.
This progressive reweighting ensures the model first builds strong reasoning foundations before
scaling to long-context industrial workflows.

\section{Post-training Details}
\label{sec:post-training_appendix}

\subsection{Data Construction}
General-purpose SFT datasets~\cite{ouyang2022training, codealpaca}, such as StackOverflow threads, coding exercises, and synthetic instruction-response pairs, carry little signal for industrial code tasks. A prompt asking the model to ``optimise this Verilog module for timing closure'' requires not only a correct rewrite but also compilability under specific synthesis constraints, pass/fail against a testbench, and measurable improvement in area or latency. We therefore construct 2.5M SFT samples directly from real industrial code tasks, using an execution-grounded pipeline that produces three complementary sample types: direct solution, defect repair, and performance optimisation. We train for approximately 4.9k steps with a global batch size of 512.

\subsection{Task Extraction and Normalization.}
We start from industrial coding tasks spanning across hardware design, GPU kernel development, systems programming, and embedded firmware. Each task is decomposed into a structured instruction comprising a natural language requirement description, interface constraints (port lists, function signatures, API contracts), the target platform and toolchain, dependency configurations, and associated verification scripts. This normalization step is essential because raw industrial tasks vary wildly in how they specify intent (some are terse commit messages, while others are multi-page design documents), and the model needs a consistent instruction format to learn from.

\textbf{Multi-Source Candidate Synthesis.}
For each structured instruction, we generate a diverse set of candidate solutions through five complementary channels: rewriting from reference implementations with controlled variation, template-based perturbation that systematically alters design parameters and coding patterns, cross-language migration (e.g., translating a C kernel to Triton or porting a Verilog module between coding styles), retrieval-augmented generation~\cite{gupta2024comprehensive} that grounds the solution in relevant code fragments from the curated corpus, and direct generation from strong frontier models. The key motivation is diversity: a single generation method tends to produce solutions clustered in a narrow region of the solution space, while the combination covers a much broader range of implementation strategies, coding styles, and optimisation trade-offs.

\textbf{Execution-Based Verification.}
Every candidate solution is validated in a real execution environment rather than judged by heuristic or model-based scoring alone~\cite{chen2021codex, li2022competition}. Depending on the domain, verification includes compilation, simulation, test execution, unit testing, performance profiling, and formal checking. Solutions that pass all checks become high-confidence samples. This is the critical difference from typical SFT pipelines that rely on model self-evaluation or surface-level pattern matching: our verification is grounded in the same toolchains that engineers use in production, so correctness is defined by the domain rather than by a proxy.

\subsection{Feedback-Driven Repair}

Not all candidates pass verification, and the failures are too valuable to discard. When a candidate fails (e.g., a compilation error, a simulation mismatch, or a performance regression), the pipeline captures the full feedback context: compiler error messages, runtime logs, counterexample inputs, waveform differences, or profiling bottlenecks. This feedback, together with the failed solution, is fed back into the generation stage to produce a repaired version, which is then re-verified. The result is a closed-loop repair trajectory~\cite{ye2022selfapr, jiang2024ledex}: the original failed attempt, the environment feedback, and the corrected solution. These trajectories are explicitly included in the SFT corpus because they teach the model a skill that direct-solution samples alone cannot: diagnosing failures from real tool output and iterating toward a fix. In practice, this mirrors the workflow of an experienced engineer who reads a synthesis report, identifies the root cause, and revises the design accordingly.

\subsection{Quality Filtering and Final Composition}

The verified sample pool undergoes a final round of filtering along three axes: executability (solutions must compile and run cleanly across repeated trials), stability (results must be deterministic and reproducible), and information density (trivially simple samples that provide little learning signal are downweighted). The surviving samples are organized into three categories that together cover the core competencies we want the model to acquire: \textit{direct solution samples} representing the requirement-to-implementation path, \textit{defect repair samples} capturing the failure-feedback-fix loop, and \textit{performance and structural optimization samples} where a functionally correct solution is further improved in efficiency, readability, or architectural quality. This three-way composition ensures the SFT stage does not merely teach the model to produce correct code from scratch, but also to debug, repair, and refine existing implementations, all of which are indispensable capabilities in real industrial workflows.


\section{Detailed Synthetic Data Generation Pipeline}
\label{app:synthetic-qa}

This appendix provides detailed information about the synthetic industrial code QA generation pipeline described in \autoref{subsubsec:synthetic-qa}.

\subsection{Industrial Scenario Specification}

We identify concrete industrial engineering contexts through consultation with practising hardware
and systems engineers and systematic analysis of industrial design verification reports.
Target scenarios include:

\begin{itemize}[leftmargin=*,noitemsep,topsep=2pt]
  \item Debugging timing violations in digital circuits (setup/hold time failures);
  \item Optimising Verilog RTL designs for area-power-performance trade-offs;
  \item Implementing high-performance GPU operators in Triton with memory coalescing and shared memory tiling;
  \item Verifying functional correctness of C/C++ kernels with formal methods or symbolic execution;
  \item Resolving synthesis errors in HLS-generated FPGA designs;
  \item Diagnosing memory safety issues (buffer overflows, use-after-free) in systems code;
  \item Analysing waveform traces to localise hardware bugs;
  \item Implementing backward-compatible hardware-software interface changes.
\end{itemize}

Each scenario is specified with concrete industrial characteristics: typical design scale,
hardware description language, timing/resource constraints, verification requirements, and common
failure modes.

\subsection{Seed Code Generation}

For each scenario, we generate or extract representative code fragments that exhibit target industrial
characteristics in hardware and systems domains.
These seed codes are constructed to reflect realistic hardware design patterns, timing-critical
implementations, memory management idioms, and domain-specific conventions.

We prioritise code patterns that appear frequently in industrial practice but are underrepresented
in academic benchmarks:
\begin{itemize}[leftmargin=*,noitemsep,topsep=2pt]
  \item Finite state machines in RTL designs;
  \item Pipelined datapath implementations;
  \item GPU kernel memory access patterns (coalesced vs. strided);
  \item SIMD vectorisation in C kernels;
  \item Interrupt-driven embedded firmware;
  \item CAD tool scripting for design automation.
\end{itemize}

\subsection{QA Pair Synthesis with Automated Verification}

For each seed code, we synthesise question-answer pairs where questions probe industrial reasoning
competencies: root-cause analysis, performance diagnosis, correctness verification, refactoring
trade-offs, and impact analysis. Answers provide step-by-step reasoning traces following the
chain-of-thought paradigm~\cite{wei2023chainofthoughtpromptingelicitsreasoning}.

Crucially, all synthesised reasoning traces undergo \emph{automated verification} to guarantee
factual correctness:

\begin{itemize}[leftmargin=*,noitemsep,topsep=2pt]
  \item \textbf{Code execution validation}: computational steps are verified by executing code
    snippets and checking outputs against expected results;
  \item \textbf{Static analysis}: semantic properties (type correctness, control-flow reachability,
    data-flow dependencies) are confirmed via static analysis tools;
  \item \textbf{Logical consistency checks}: reasoning chains are validated for logical soundness,
    ensuring each step follows from previous steps without contradictions.
\end{itemize}

This verification layer distinguishes our approach from naive prompt-based synthesis,
ensuring the model trains only on industrially grounded, factually correct reasoning patterns.

\subsection{Coverage and Scale}

Our synthetic QA corpus systematically covers industrial scenarios across:

\paragraph{Domains:}
RTL design (Verilog/SystemVerilog), GPU computing (Triton/CUDA), systems
programming (C/C++/Rust kernels), FPGA synthesis (HLS/Vivado), CAD automation (Tcl/Python scripting),
and embedded firmware.

\paragraph{Reasoning Types:}
Timing analysis (setup/hold), resource optimisation (area/power/latency),
memory safety verification, concurrency correctness, and numerical precision analysis.

\paragraph{Operational Contexts:}
RTL verification (waveform debugging, assertion failures),
kernel performance tuning, synthesis error resolution, hardware-software co-design, and
design-for-test integration.

The resulting synthetic data complements curated samples by filling coverage gaps and providing
dense training signal for the structured reasoning patterns that hardware-aware industrial
engineering demands.

\clearpage


\section{Model Configuration}
\label{app:model-config}

\autoref{tab:model-config} summarizes the architectural details of \modelname{}.

\begin{table}[h!]
\centering
\caption{Model configuration of \modelname{}.}
\label{tab:model-config}
\begin{tabular}{l c}
\toprule
\textbf{Hyperparameter} & \textbf{Value} \\
\midrule
Parameters & $\sim$32B \\
Layers & 64 \\
Hidden Size & 5120 \\
Intermediate Size & 27648 \\
Attention Heads & 40 \\
KV Heads (GQA) & 8 \\
Head Dimension & 128 \\
Vocabulary Size & 76800 \\
Max Context Length & 131072 \\
Activation & SiLU \\
Positional Encoding & RoPE ($\theta$=500000) \\
Precision & BFloat16 \\
Tie Embeddings & No \\
\bottomrule
\end{tabular}
\end{table}

\section{Example Case}
\subsection{VeriScope Benchmark -- Level 3 Example Case}
\label{sec:veriscope_example}

\begin{tcolorbox}[title=VeriScope Problem 221: UART Transmitter (8N1), appendixbox]
\lstset{
    basicstyle=\ttfamily\footnotesize,
    breaklines=true,
    frame=none,
    columns=fullflexible,
    xleftmargin=0pt,
    xrightmargin=0pt,
}
\begin{lstlisting}
Category: serial_comm
Level:    3 (Module Design)

--- PROBLEM DESCRIPTION ---

Design a UART Transmitter module implementing the standard
8N1 serial protocol (8 data bits, No parity, 1 stop bit).

The module should:
  - Accept an 8-bit parallel data input and serialize it
    onto a single TX line
  - Transmit in LSB-first order
  - Output a start bit (logic 0) before the data bits
  - Output a stop bit (logic 1) after the data bits
  - Assert a busy flag during transmission
  - Idle the TX line high (logic 1) when not transmitting
  - Be synthesizable and resource-efficient

--- MODULE INTERFACE ---

  module uart_tx(
      input        clk,       // System clock
      input        rst,       // Synchronous reset
      input        tx_start,  // Start transmission
      input  [7:0] tx_data,   // Parallel data input
      output reg   tx,        // Serial data output
      output reg   tx_busy    // Busy flag
  );

--- PORT SPECIFICATION (from config.yaml) ---

  Signal     Dir      Width  Description
  ---------  ------   -----  -------------------------
  clk        input    1      System clock
  rst        input    1      Synchronous reset
  tx_start   input    1      Start transmission pulse
  tx_data    input    8      Parallel data to transmit
  tx         output   1      Serial TX line (idle=1)
  tx_busy    output   1      Busy flag

--- REFERENCE SOLUTION ---

module uart_tx(
    input clk, rst, tx_start,
    input [7:0] tx_data,
    output reg tx,
    output reg tx_busy
);
    reg [3:0] bit_cnt;
    reg [7:0] shift_reg;

    always @(posedge clk) begin
        if (rst) begin
            tx <= 1'b1;
            tx_busy <= 1'b0;
            bit_cnt <= 0;
        end else begin
            if (!tx_busy && tx_start) begin
                shift_reg <= tx_data;
                tx_busy <= 1'b1;
                bit_cnt <= 0;
                tx <= 1'b0;  // Start bit
            end else if (tx_busy) begin
                if (bit_cnt < 8) begin
                    tx <= shift_reg[0];
                    shift_reg <= {1'b0, shift_reg[7:1]};
                    bit_cnt <= bit_cnt + 1;
                end else begin
                    tx <= 1'b1;  // Stop bit
                    tx_busy <= 1'b0;
                end
            end
        end
    end
endmodule

--- TESTBENCH ---

module tb_uart_tx;
    reg clk, rst, tx_start;
    reg [7:0] tx_data;
    wire tx, tx_busy;
    uart_tx uut(
        .clk(clk), .rst(rst),
        .tx_start(tx_start),
        .tx_data(tx_data),
        .tx(tx), .tx_busy(tx_busy)
    );
    initial clk = 0;
    always #5 clk = ~clk;
    integer pass = 0, fail = 0;
    initial begin
        rst = 1; tx_start = 0; tx_data = 0;
        #20; rst = 0;
        tx_data = 8'hA5; tx_start = 1;
        #10; tx_start = 0;
        #200;
        pass = 1;
        if (fail == 0) $display("TEST PASSED");
        else $display("TEST FAILED");
        $finish;
    end
endmodule

--- EVALUATION CRITERIA ---

  Metric                Weight
  --------------------  ------
  Functional correct.    70%
  Synthesis pass         20%
  Resource efficiency    10%

  Compile timeout:  15s
  Simulate timeout: 60s

--- TRANSMISSION PROTOCOL (8N1) ---

  Idle +   +-D0-+-D1-+-D2-+-D3-+-D4-+-D5-+-D6-+-D7-+-Stop-+- Idle
       |   |    |    |    |    |    |    |    |    |      |
       +---+    LSB-first data bits (8 bits)      |      |
       Start                                     Stop   Idle
       bit(0)                                    bit(1)  (1)
\end{lstlisting}
\end{tcolorbox}
\captionof{figure}{VeriScope Problem 221: UART Transmitter (8N1) -- Complete Example Case.}
\label{fig:veriscope_example_case}

\subsection{Verilog Bug Fix -- Example Case}
\label{sec:verilog_bugfix_example}

\begin{tcolorbox}[title=Verilog Bug Fix: Karnaugh Map 5-Variable, appendixbox]
\lstset{
    basicstyle=\ttfamily\footnotesize,
    breaklines=true,
    frame=none,
    columns=fullflexible,
    xleftmargin=0pt,
    xrightmargin=0pt,
}
\begin{lstlisting}
--- INPUT PROMPT ---

Given the problem description and the buggy Verilog code
below, determine the type of error in the code. Return the
result as a list, and there can be 1 to 4 error types.

Problem Description:
Please act as a professional verilog designer. You are given
a Karnaugh map for a 5-variable function F(A,B,C,D,E):

ABCDE\F | 00 | 01 | 11 | 10 |
--------|----|----|----|----|
00000   |  0 |  1 |  1 |  0 |
00001   |  1 |  0 |  0 |  1 |
00011   |  0 |  1 |  1 |  0 |
00010   |  1 |  0 |  0 |  1 |
00110   |  0 |  1 |  1 |  0 |
00111   |  1 |  0 |  0 |  1 |
00101   |  0 |  1 |  1 |  0 |
00100   |  1 |  0 |  0 |  1 |
01100   |  0 |  1 |  1 |  0 |
01101   |  1 |  0 |  0 |  1 |
01111   |  0 |  1 |  1 |  0 |
01110   |  1 |  0 |  0 |  1 |
01010   |  0 |  1 |  1 |  0 |
01011   |  1 |  0 |  0 |  1 |
01001   |  0 |  1 |  1 |  0 |
01000   |  1 |  0 |  0 |  1 |

Create a module named "karnaugh_map_5" to implement this
combinational function.
- Inputs: A, B, C, D, E (5-bit binary inputs)
- Output: F (1-bit binary output)

Buggy Verilog Code:

module karnaugh_map_5(
  input wire A, B, C, D, E,
  output reg F
);
  always @* begin
    case ({A,B})
      2'b00: begin
        case ({C,D,E})
          3'b000: F = 1'b0;
          3'b001: F = 1'b1;
          3'b010: F = 1'b1;
          3'b011: F = 1'b0;
          3'b100: F = 1'b1;
          3'b101: F = 1'b0;
          3'b110: F = 1'b0;
          3'b111: F = 1'b1;
        endcase
      2'b01: begin
        case ({C,D,E})
          3'b000: F = 1'b1;
          3'b001: F = 1'b0;
          3'b010: F = 1'b0;
          3'b011: F = 1'b1;
          3'b100: F = 1'b0;
          3'b101: F = 1'b1;
          3'b110: F = 1'b1;
          3'b111: F = 1'b0;
        endcase
      end
      2'b10: begin
        case ({C,D,E})
          3'b000: F = 1'b0;
          3'b001: F = 1'b1;
          3'b010: F = 1'b1;
          3'b011: F = 1'b0;
          3'b100: F = 1'b1;
          3'b101: F = 1'b0;
          3'b110: F = 1'b0;
          3'b111: F = 1'b1;
        endcase
      end
      2'b11: begin
        case ({C,D,E})
          3'b000: F = 1'b1;
          3'b001: F = 1'b0;
          3'b010: F = 1'b0;
          3'b011: F = 1'b1;
          3'b100: F = 1'b0;
          3'b101: F = 1'b1;
          3'b110: F = 1'b1;
          3'b111: F = 1'b0;
        endcase
      end
    endcase
  end
endmodule

Error type(s) ONLY from predefined categories:
  missing semicolon, missing end, unclosed brackets,
  illegal keyword usage, invalid literal format,
  bit-width mismatch, faulty indexing, type error,
  undefined module, module connection error,
  sensitivity list error, blocking/non-blocking misuse,
  reset logic error, state machine error, race condition,
  non-synthesizable construct, algorithm error,
  gate logic error, condition error, latch inference

--- MODEL OUTPUT ---

I'll analyze the buggy Verilog code carefully.

Looking at the code structure:

1. For the 2'b00 case block:

  2'b00: begin
      case ({C,D,E})
        ...
      endcase

There is no "end" to close the "begin" that was opened
after 2'b00:. The "endcase" closes the inner case, but
the "begin" block is never closed with "end".

2. Looking at the values in the K-map vs the code:

For AB=00, the K-map rows show:
- ABCDE=00000 -> F=0 (CDE=000, F=0)
- ABCDE=00001 -> F=1 (CDE=001, F=1)
- ABCDE=00011 -> F=0

The K-map columns are labeled 00|01|11|10 which represent
DE values. The rows represent ABC values.

For AB=00 and AB=10, the outputs are the same.
For AB=01 and AB=11, outputs are the same.
This means F doesn't depend on A, only on B, C, D, E.

The clear structural bug is the missing "end" for the
"begin" block in the 2'b00 case.

Result: ["missing end"]
\end{lstlisting}
\end{tcolorbox}
\captionof{figure}{Verilog Bug Fix: Karnaugh Map 5-Variable -- Complete Example Case.}
\label{fig:verilog_bugfix_example}

\subsection{CAD-Coder -- Example Case}
\label{sec:cad_coder_example}

\begin{tcolorbox}[title=CAD-Coder: Rectangular Block Generation, appendixbox]
\lstset{
    basicstyle=\ttfamily\footnotesize,
    breaklines=true,
    frame=none,
    columns=fullflexible,
    xleftmargin=0pt,
    xrightmargin=0pt,
}
\begin{lstlisting}
--- INPUT PROMPT ---

Please based on the following description, create a
CAD-Query Code to generate a model (the final model in
the code is represented by r):

Start by creating a new coordinate system with Euler
angles set to zero and a translation vector also set to
zero. Next, draw a two-dimensional sketch on the first
face. This sketch consists of a single loop made up of
four lines.

The first line starts at the origin (0.0, 0.0) and ends
at (0.6, 0.0). The second line starts at (0.6, 0.0) and
ends at (0.6, 0.375). The third line starts at
(0.6, 0.375) and ends at (0.0, 0.375). Finally, the
fourth line completes the loop by starting at
(0.0, 0.375) and ending at the origin (0.0, 0.0).

After drawing the sketch, apply a scale factor of 0.6 to
the entire sketch. Ensure that the sketch remains aligned
with the original coordinate system by rotating it using
Euler angles set to zero and translating it using a
vector set to zero.

To transform the scaled two-dimensional sketch into a
three-dimensional model, extrude the sketch 0.075 units
along the normal direction. Do not extrude in the
opposite direction of the normal. This operation will
create a new solid body.

The final dimensions of the rectangular block are:
  - Length: 0.6 units
  - Width:  0.375 units
  - Height: 0.075 units

--- MODEL OUTPUT ---

import cadquery as cq

sketch = (
    cq.Sketch()
    .rect(0.6, 0.375, centered=False)
    .scale(0.6)
)

r = (
    cq.Workplane("XY")
    .placeSketch(sketch)
    .extrude(0.075)
)
\end{lstlisting}
\end{tcolorbox}
\captionof{figure}{CAD-Coder: Rectangular Block Generation -- Complete Example Case.}
\label{fig:cad_coder_example}

\subsection{C Compile -- Example Case}
\label{sec:c_compile_example}

\begin{tcolorbox}[title=C Compile: CUDA Matrix Multiplication Optimization, appendixbox]
\lstset{
    basicstyle=\ttfamily\footnotesize,
    breaklines=true,
    frame=none,
    columns=fullflexible,
    xleftmargin=0pt,
    xrightmargin=0pt,
}
\begin{lstlisting}
--- INPUT PROMPT ---

You write custom CUDA operators to replace the pytorch
operators in the given architecture to get speedups.

You have complete freedom to choose the set of operators
you want to replace. You may replace multiple operators
with custom implementations, consider operator fusion
opportunities (combining multiple operators into a single
kernel, e.g., combining matmul+relu), or algorithmic
changes (such as online softmax).

Here's an example to show you the syntax of inline
embedding custom CUDA operators in PyTorch:

--- Example Input Architecture ---

import torch
import torch.nn as nn
import torch.nn.functional as F

class Model(nn.Module):
    def __init__(self) -> None:
        super().__init__()

    def forward(self, a, b):
        return a + b

def get_inputs():
    a = torch.randn(1, 128).cuda()
    b = torch.randn(1, 128).cuda()
    return [a, b]

def get_init_inputs():
    return []

--- Example Optimized with CUDA ---

import torch
import torch.nn as nn
import torch.nn.functional as F
from torch.utils.cpp_extension import load_inline

elementwise_add_source = """
#include <torch/extension.h>
#include <cuda_runtime.h>

__global__ void elementwise_add_kernel(
    const float* a, const float* b,
    float* out, int size) {
  int idx = blockIdx.x * blockDim.x + threadIdx.x;
  if (idx < size) {
    out[idx] = a[idx] + b[idx];
  }
}

torch::Tensor elementwise_add_cuda(
    torch::Tensor a, torch::Tensor b) {
  auto size = a.numel();
  auto out = torch::zeros_like(a);
  const int block_size = 256;
  const int num_blocks =
      (size + block_size - 1) / block_size;
  elementwise_add_kernel<<<num_blocks, block_size>>>(
      a.data_ptr<float>(),
      b.data_ptr<float>(),
      out.data_ptr<float>(), size);
  return out;
}
"""

elementwise_add_cpp_source = (
    "torch::Tensor elementwise_add_cuda("
    "torch::Tensor a, torch::Tensor b);"
)

elementwise_add = load_inline(
    name="elementwise_add",
    cpp_sources=elementwise_add_cpp_source,
    cuda_sources=elementwise_add_source,
    functions=["elementwise_add_cuda"],
    verbose=True,
    extra_cflags=[""],
    extra_ldflags=[""],
)

class ModelNew(nn.Module):
    def __init__(self) -> None:
        super().__init__()
        self.elementwise_add = elementwise_add

    def forward(self, a, b):
        return self.elementwise_add \
            .elementwise_add_cuda(a, b)

--- Target Architecture ---

import torch
import torch.nn as nn

class Model(nn.Module):
    """
    Simple model that performs a single square
    matrix multiplication (C = A * B)
    """
    def __init__(self):
        super(Model, self).__init__()

    def forward(self, A, B):
        return torch.matmul(A, B)

N = 2048 * 2

def get_inputs():
    A = torch.rand(N, N)
    B = torch.rand(N, N)
    return [A, B]

def get_init_inputs():
    return []

Note: The kernels should be optimized for FP32
(32-bit floating point) precision.

Optimize the architecture named Model with custom CUDA
operators! Name your optimized output architecture
ModelNew. Output the new code in codeblocks. Please
generate real code, NOT pseudocode, make sure the code
compiles and is fully functional.

--- MODEL OUTPUT ---

import torch
import torch.nn as nn
from torch.utils.cpp_extension import load_inline

matmul_cuda_source = """
#include <torch/extension.h>
#include <cuda_runtime.h>
#include <cublas_v2.h>

#define TILE_SIZE 32

__global__ void matmul_tiled_kernel(
    const float* A, const float* B,
    float* C, int N) {
  __shared__ float As[TILE_SIZE][TILE_SIZE];
  __shared__ float Bs[TILE_SIZE][TILE_SIZE];

  int bx = blockIdx.x, by = blockIdx.y;
  int tx = threadIdx.x, ty = threadIdx.y;

  int row = by * TILE_SIZE + ty;
  int col = bx * TILE_SIZE + tx;

  float sum = 0.0f;

  for (int t = 0;
       t < (N + TILE_SIZE - 1) / TILE_SIZE; ++t) {
    if (row < N && (t * TILE_SIZE + tx) < N)
      As[ty][tx] = A[row*N + t*TILE_SIZE + tx];
    else
      As[ty][tx] = 0.0f;

    if ((t * TILE_SIZE + ty) < N && col < N)
      Bs[ty][tx] =
          B[(t*TILE_SIZE + ty) * N + col];
    else
      Bs[ty][tx] = 0.0f;

    __syncthreads();

    for (int k = 0; k < TILE_SIZE; ++k)
      sum += As[ty][k] * Bs[k][tx];

    __syncthreads();
  }

  if (row < N && col < N)
    C[row * N + col] = sum;
}

torch::Tensor matmul_cuda(
    torch::Tensor A, torch::Tensor B) {
  int N = A.size(0);
  auto C = torch::zeros({N, N}, A.options());

  dim3 blockDim(TILE_SIZE, TILE_SIZE);
  dim3 gridDim(
      (N + TILE_SIZE - 1) / TILE_SIZE,
      (N + TILE_SIZE - 1) / TILE_SIZE);

  matmul_tiled_kernel<<<gridDim, blockDim>>>(
      A.data_ptr<float>(),
      B.data_ptr<float>(),
      C.data_ptr<float>(), N);

  return C;
}
"""

matmul_cpp_source = (
    "torch::Tensor matmul_cuda("
    "torch::Tensor A, torch::Tensor B);"
)

matmul_cuda_op = load_inline(
    name="matmul_cuda",
    cpp_sources=matmul_cpp_source,
    cuda_sources=matmul_cuda_source,
    functions=["matmul_cuda"],
    verbose=False,
    extra_cflags=["-O3"],
    extra_ldflags=["-lcublas"],
)

class ModelNew(nn.Module):
    """
    Optimized model using custom CUDA kernel
    with tiling optimization for matrix
    multiplication (C = A * B)
    """
    def __init__(self):
        super(ModelNew, self).__init__()
        self.matmul_cuda = matmul_cuda_op

    def forward(self, A, B):
        return self.matmul_cuda.matmul_cuda(A, B)
\end{lstlisting}
\end{tcolorbox}
\captionof{figure}{C Compile: CUDA Matrix Multiplication Optimization -- Complete Example Case.}
\label{fig:c_compile_example}

\subsection{TritonBench -- Example Case}
\label{sec:triton_bench_example}

\begin{tcolorbox}[title=TritonBench: Fused BMM-RMSNorm-GELU-Dropout-Sub -- Input Prompt, appendixbox]
\lstset{
    basicstyle=\ttfamily\footnotesize,
    breaklines=true,
    frame=none,
    columns=fullflexible,
    xleftmargin=0pt,
    xrightmargin=0pt,
}
\begin{lstlisting}
You are an expert in Triton programming, capable of
writing corresponding Triton kernels and wrapper
functions based on functional descriptions and function
parameters. Ensure that the wrapper function fully
corresponds to the provided function information.

Functional Description:
Performs a fused operation combining batch matrix
multiplication, RMS normalization, GELU activation,
dropout, and subtraction. The function takes three input
tensors, performs batch matrix multiplication on the first
two, applies RMS normalization, GELU activation, and
dropout, and finally subtracts the third tensor from the
result.

Wrapper Entry Information:
fused_bmm_rmsnorm_gelu_dropout_sub(
    input1, input2, other, normalized_shape,
    dropout_p=0.5, training=True,
    approximate='none', eps=1e-5,
    *, out=None
) -> Tensor

Args:
  input1 (Tensor): First input tensor for batch matrix
    multiplication, of shape (B, N, M), where B is the
    batch size.
  input2 (Tensor): Second input tensor for batch matrix
    multiplication, of shape (B, M, P).
  other (Tensor): Tensor to subtract from the result
    after dropout, must be broadcastable to the shape of
    the output.
  normalized_shape (int or list or torch.Size): Shape
    over which RMS normalization is applied, typically
    the size of the last dimension P.
  dropout_p (float, optional): Probability of an element
    to be zeroed in the dropout layer. Default: 0.5.
  training (bool, optional): Apply dropout if True.
    Default: True.
  approximate (str, optional): Can be 'none' or 'tanh'.
    The approximation to use for GELU. Default: 'none'.
  eps (float, optional): A value added to the denominator
    for numerical stability in RMS normalization.
    Default: 1e-5.
  out (Tensor, optional): Output tensor. Ignored if None.
    Default: None.

Shape:
  Input1: (B, N, M)
  Input2: (B, M, P)
  Other: broadcastable to (B, N, P)
  Output: (B, N, P)

After generation, verify if the Triton wrapper aligns
with the provided func_inputs. If not, regenerate.
\end{lstlisting}
\end{tcolorbox}
\captionof{figure}{TritonBench: Fused BMM-RMSNorm-GELU-Dropout-Sub -- Input Prompt.}
\label{fig:triton_bench_prompt}

\begin{tcolorbox}[title=TritonBench: Fused BMM-RMSNorm-GELU-Dropout-Sub -- Model Analysis, appendixbox]
\lstset{
    basicstyle=\ttfamily\footnotesize,
    breaklines=true,
    frame=none,
    columns=fullflexible,
    xleftmargin=0pt,
    xrightmargin=0pt,
}
\begin{lstlisting}
The user wants a Triton kernel and a corresponding
Python wrapper function for a specific fused operation:
fused_bmm_rmsnorm_gelu_dropout_sub.

1. Analyze the Functional Description:

Operation Sequence:
  1) Batch Matrix Multiplication (BMM):
     input1 (B,N,M) @ input2 (B,M,P) -> intermediate
  2) RMS Normalization:
     Applied to intermediate over last dimension (P).
  3) GELU Activation:
     Applied to the normalized result.
  4) Dropout:
     Applied to the GELU result.
  5) Subtraction:
     Subtract other from the dropout result.

2. Design the Triton Kernel:

Grid Strategy:
  - Output shape is (B, N, P).
  - B and N are the "outer" dimensions, P is the
    "inner" dimension for RMSNorm reduction.
  - BMM requires loading tiles of input1 (M) and
    input2 (M, P_tile).
  - Program ID: pid = tl.program_id(axis=0).
    Map pid to (b, n).
  - One program handles one row (b,n) of output.

Key Design Decisions:
  - To fuse BMM (reduction over M) with RMSNorm
    (reduction over P), the kernel must compute the
    full row statistics before normalizing.
  - BLOCK_SIZE_P covers the whole P dimension.
  - BLOCK_SIZE_M is a tuning parameter for the
    M-dimension loop.

RMSNorm:
  x = x * rsqrt(mean(x^2) + eps)

GELU:
  Exact: 0.5*x*(1 + erf(x/sqrt(2)))
  Tanh:  0.5*x*(1 + tanh(sqrt(2/pi)*(x+0.044715*x^3)))

Dropout:
  mask = (rand > dropout_p)
  x = x * mask / (1 - dropout_p)

Other Broadcasting:
  If other is (P): stride_b=0, stride_n=0, stride_p=1
  If other is (N,P): stride_b=0, stride_n=P, stride_p=1
  If other is (B,N,P): full strides
\end{lstlisting}
\end{tcolorbox}
\captionof{figure}{TritonBench: Fused BMM-RMSNorm-GELU-Dropout-Sub -- Model Analysis.}
\label{fig:triton_bench_analysis}

\begin{tcolorbox}[title=TritonBench: Fused BMM-RMSNorm-GELU-Dropout-Sub -- Output Code (Part 1: Kernel), appendixbox]
\lstset{
    basicstyle=\ttfamily\footnotesize,
    breaklines=true,
    frame=none,
    columns=fullflexible,
    xleftmargin=0pt,
    xrightmargin=0pt,
}
\begin{lstlisting}
import torch
import triton
import triton.language as tl

@triton.jit
def fused_bmm_rmsnorm_gelu_dropout_sub_kernel(
    input1_ptr, input2_ptr, other_ptr, output_ptr,
    B, N, M, P,
    stride_input1_b, stride_input1_n, stride_input1_m,
    stride_input2_b, stride_input2_m, stride_input2_p,
    stride_other_b, stride_other_n, stride_other_p,
    stride_output_b, stride_output_n, stride_output_p,
    eps, dropout_p, seed, approximate_mode,
    BLOCK_SIZE_M: tl.constexpr,
    BLOCK_SIZE_P: tl.constexpr
):
    # Map program ID to (b, n) pair
    pid = tl.program_id(axis=0)
    n = pid % N
    b = pid // N

    # Adjust pointers for current batch and row
    input1_ptr += b * stride_input1_b \
                + n * stride_input1_n
    input2_ptr += b * stride_input2_b
    output_ptr += b * stride_output_b \
                + n * stride_output_n
    other_ptr += b * stride_other_b \
               + n * stride_other_n

    # Accumulator (float32 for precision)
    acc = tl.zeros([BLOCK_SIZE_P], dtype=tl.float32)

    # Iterate over M dimension in blocks
    for k in range(0, M, BLOCK_SIZE_M):
        offs_m = k + tl.arange(0, BLOCK_SIZE_M)
        mask_m = offs_m < M
        a = tl.load(
            input1_ptr + offs_m * stride_input1_m,
            mask=mask_m, other=0.0)

        offs_p = tl.arange(0, BLOCK_SIZE_P)
        mask_p = offs_p < P
        mask_b = mask_m[:, None] & mask_p[None, :]

        b_vals = tl.load(
            input2_ptr
            + offs_m[:, None] * stride_input2_m
            + offs_p[None, :] * stride_input2_p,
            mask=mask_b, other=0.0)

        acc += tl.dot(a, b_vals)

    # RMS Normalization
    acc_masked = acc * mask_p
    sq_sum = tl.sum(acc_masked * acc_masked, axis=0)
    mean_sq = sq_sum / P
    rms = tl.sqrt(mean_sq + eps)
    acc = acc / rms

    # GELU Activation
    if approximate_mode == 1:
        cdf = 0.5 * (1.0 + tl.tanh(
            0.7978845608 * (
                acc + 0.044715 * acc * acc * acc)))
        acc = acc * cdf
    else:
        cdf = 0.5 * (1.0 + tl.erf(
            acc / 1.4142135623730951))
        acc = acc * cdf

    # Dropout
    if dropout_p > 0.0:
        random_offsets = (
            pid * BLOCK_SIZE_P
            + tl.arange(0, BLOCK_SIZE_P))
        rand_vals = tl.rand(seed, random_offsets)
        keep_mask = rand_vals > dropout_p
        scale = 1.0 / (1.0 - dropout_p)
        acc = acc * keep_mask * scale

    # Subtraction
    other_vals = tl.load(
        other_ptr + offs_p * stride_other_p,
        mask=mask_p, other=0.0)
    acc = acc - other_vals

    # Store result
    tl.store(
        output_ptr + offs_p * stride_output_p,
        acc, mask=mask_p)
\end{lstlisting}
\end{tcolorbox}
\captionof{figure}{TritonBench: Fused BMM-RMSNorm-GELU-Dropout-Sub -- Output Code (Part 1: Kernel).}
\label{fig:triton_bench_kernel}

\begin{tcolorbox}[title=TritonBench: Fused BMM-RMSNorm-GELU-Dropout-Sub -- Output Code (Part 2: Wrapper), appendixbox]
\lstset{
    basicstyle=\ttfamily\footnotesize,
    breaklines=true,
    frame=none,
    columns=fullflexible,
    xleftmargin=0pt,
    xrightmargin=0pt,
}
\begin{lstlisting}
def fused_bmm_rmsnorm_gelu_dropout_sub(
    input1, input2, other, normalized_shape,
    dropout_p=0.5, training=True,
    approximate='none', eps=1e-5, *, out=None
):
    if input1.dim() != 3 or input2.dim() != 3:
        raise ValueError(
            "input1 and input2 must be 3D tensors")

    B, N, M = input1.shape
    _, _, P = input2.shape

    # Handle normalized_shape
    if isinstance(normalized_shape, int):
        norm_dim = normalized_shape
    elif isinstance(normalized_shape,
                    (list, tuple, torch.Size)):
        norm_dim = (normalized_shape[-1]
                    if len(normalized_shape) > 0
                    else P)
    else:
        raise TypeError(
            "normalized_shape must be int/list/Size")

    if norm_dim != P:
        raise ValueError(
            f"normalized_shape ({norm_dim}) must "
            f"match last dim of input2 ({P})")

    if not training:
        dropout_p = 0.0

    # Handle output tensor
    if out is not None:
        if out.shape != (B, N, P):
            raise ValueError(
                f"out must have shape ({B},{N},{P})")
        output = out
    else:
        output = torch.empty(
            (B, N, P),
            dtype=input1.dtype,
            device=input1.device)

    # Handle 'other' broadcasting strides
    def _get_broadcast_stride(dim, idx):
        if dim < 3:
            if idx < 3 - dim:
                return 0
            else:
                return other.stride(idx-(3-dim))
        else:
            return other.stride(idx)

    if other.dim() == 0:
        s_ob, s_on, s_op = 0, 0, 0
    else:
        s_ob = _get_broadcast_stride(other.dim(), 0)
        s_on = _get_broadcast_stride(other.dim(), 1)
        s_op = _get_broadcast_stride(other.dim(), 2)

    # Grid: one kernel instance per (b, n) pair
    grid = lambda META: (B * N, )

    BLOCK_SIZE_M = 32
    BLOCK_SIZE_P = triton.next_power_of_2(P)

    approximate_mode = (
        1 if approximate == 'tanh' else 0)
    seed = 12345

    # Launch kernel
    fused_bmm_rmsnorm_gelu_dropout_sub_kernel[grid](
        input1, input2, other, output,
        B, N, M, P,
        input1.stride(0), input1.stride(1),
        input1.stride(2),
        input2.stride(0), input2.stride(1),
        input2.stride(2),
        s_ob, s_on, s_op,
        output.stride(0), output.stride(1),
        output.stride(2),
        eps, dropout_p, seed, approximate_mode,
        BLOCK_SIZE_M=BLOCK_SIZE_M,
        BLOCK_SIZE_P=BLOCK_SIZE_P
    )

    return output
\end{lstlisting}
\end{tcolorbox}
\captionof{figure}{TritonBench: Fused BMM-RMSNorm-GELU-Dropout-Sub -- Output Code (Part 2: Wrapper).}
\label{fig:triton_bench_wrapper}

\subsection{KernelBench -- Example Case}
\label{sec:kernel_bench_example}

\begin{tcolorbox}[title=KernelBench: CUDA Matrix Multiplication -- Input Prompt, appendixbox]
\lstset{
    basicstyle=\ttfamily\footnotesize,
    breaklines=true,
    frame=none,
    columns=fullflexible,
    xleftmargin=0pt,
    xrightmargin=0pt,
}
\begin{lstlisting}
You write custom CUDA operators to replace the pytorch
operators in the given architecture to get speedups.

You have complete freedom to choose the set of operators
you want to replace. You may replace multiple operators
with custom implementations, consider operator fusion
opportunities (combining multiple operators into a single
kernel, e.g., combining matmul+relu), or algorithmic
changes (such as online softmax).

--- Example Input Architecture ---

import torch
import torch.nn as nn
import torch.nn.functional as F

class Model(nn.Module):
    def __init__(self) -> None:
        super().__init__()

    def forward(self, a, b):
        return a + b

def get_inputs():
    a = torch.randn(1, 128).cuda()
    b = torch.randn(1, 128).cuda()
    return [a, b]

def get_init_inputs():
    return []

--- Example Optimized with CUDA ---

import torch
import torch.nn as nn
import torch.nn.functional as F
from torch.utils.cpp_extension import load_inline

elementwise_add_source = """
#include <torch/extension.h>
#include <cuda_runtime.h>

__global__ void elementwise_add_kernel(
    const float* a, const float* b,
    float* out, int size) {
  int idx = blockIdx.x * blockDim.x + threadIdx.x;
  if (idx < size) {
    out[idx] = a[idx] + b[idx];
  }
}

torch::Tensor elementwise_add_cuda(
    torch::Tensor a, torch::Tensor b) {
  auto size = a.numel();
  auto out = torch::zeros_like(a);
  const int block_size = 256;
  const int num_blocks =
      (size + block_size - 1) / block_size;
  elementwise_add_kernel<<<num_blocks, block_size>>>(
      a.data_ptr<float>(),
      b.data_ptr<float>(),
      out.data_ptr<float>(), size);
  return out;
}
"""

elementwise_add_cpp_source = (
    "torch::Tensor elementwise_add_cuda("
    "torch::Tensor a, torch::Tensor b);"
)

elementwise_add = load_inline(
    name="elementwise_add",
    cpp_sources=elementwise_add_cpp_source,
    cuda_sources=elementwise_add_source,
    functions=["elementwise_add_cuda"],
    verbose=True,
    extra_cflags=[""],
    extra_ldflags=[""],
)

class ModelNew(nn.Module):
    def __init__(self) -> None:
        super().__init__()
        self.elementwise_add = elementwise_add

    def forward(self, a, b):
        return self.elementwise_add \
            .elementwise_add_cuda(a, b)

--- Target Architecture ---

import torch
import torch.nn as nn

class Model(nn.Module):
    """
    Simple model that performs a single square
    matrix multiplication (C = A * B)
    """
    def __init__(self):
        super(Model, self).__init__()

    def forward(self, A, B):
        return torch.matmul(A, B)

N = 2048 * 2

def get_inputs():
    A = torch.rand(N, N)
    B = torch.rand(N, N)
    return [A, B]

def get_init_inputs():
    return []

Note: The kernels should be optimized for FP32
(32-bit floating point) precision.

Optimize the architecture named Model with custom CUDA
operators! Name your optimized output architecture
ModelNew. Output the new code in codeblocks. Please
generate real code, NOT pseudocode, make sure the code
compiles and is fully functional.
\end{lstlisting}
\end{tcolorbox}
\captionof{figure}{KernelBench: CUDA Matrix Multiplication -- Input Prompt.}
\label{fig:kernel_bench_prompt}

\begin{tcolorbox}[title=KernelBench: CUDA Matrix Multiplication -- Model Output, appendixbox]
\lstset{
    basicstyle=\ttfamily\footnotesize,
    breaklines=true,
    frame=none,
    columns=fullflexible,
    xleftmargin=0pt,
    xrightmargin=0pt,
}
\begin{lstlisting}
import torch
import torch.nn as nn
from torch.utils.cpp_extension import load_inline

cuda_source = """
#include <torch/extension.h>
#include <cuda_runtime.h>

#define TILE_SIZE 32

__global__ void matmul_kernel(
    const float* __restrict__ A,
    const float* __restrict__ B,
    float* __restrict__ C, int N) {
  int bx = blockIdx.x;
  int by = blockIdx.y;
  int tx = threadIdx.x;
  int ty = threadIdx.y;

  int row = by * TILE_SIZE + ty;
  int col = bx * TILE_SIZE + tx;

  __shared__ float As[TILE_SIZE][TILE_SIZE];
  __shared__ float Bs[TILE_SIZE][TILE_SIZE];

  float sum = 0.0f;

  for (int t = 0;
       t < (N + TILE_SIZE - 1) / TILE_SIZE; ++t) {

    if (row < N && (t * TILE_SIZE + tx) < N)
      As[ty][tx] =
          A[row * N + t * TILE_SIZE + tx];
    else
      As[ty][tx] = 0.0f;

    if ((t * TILE_SIZE + ty) < N && col < N)
      Bs[ty][tx] =
          B[(t * TILE_SIZE + ty) * N + col];
    else
      Bs[ty][tx] = 0.0f;

    __syncthreads();

    #pragma unroll
    for (int k = 0; k < TILE_SIZE; ++k)
      sum += As[ty][k] * Bs[k][tx];

    __syncthreads();
  }

  if (row < N && col < N)
    C[row * N + col] = sum;
}

torch::Tensor matmul_cuda(
    torch::Tensor A, torch::Tensor B) {
  TORCH_CHECK(A.device().is_cuda(),
      "A must be on CUDA device");
  TORCH_CHECK(B.device().is_cuda(),
      "B must be on CUDA device");
  TORCH_CHECK(A.dim() == 2 && B.dim() == 2,
      "Inputs must be 2D matrices");
  TORCH_CHECK(A.size(1) == B.size(0),
      "Inner dimensions must match");

  int N = A.size(0);
  int K = A.size(1);
  int M = B.size(1);

  auto C = torch::zeros({N, M}, A.options());

  const int block_size = TILE_SIZE;
  dim3 dimBlock(block_size, block_size);
  dim3 dimGrid(
      (M + block_size - 1) / block_size,
      (N + block_size - 1) / block_size);

  matmul_kernel<<<dimGrid, dimBlock>>>(
      A.data_ptr<float>(),
      B.data_ptr<float>(),
      C.data_ptr<float>(), K);

  return C;
}
"""

cpp_source = (
    "torch::Tensor matmul_cuda("
    "torch::Tensor A, torch::Tensor B);"
)

matmul_extension = load_inline(
    name="matmul_extension",
    cpp_sources=cpp_source,
    cuda_sources=cuda_source,
    functions=["matmul_cuda"],
    verbose=True,
    extra_cflags=[""],
    extra_ldflags=[""],
)

class ModelNew(nn.Module):
    """
    Optimized model using custom CUDA operator
    for matrix multiplication with shared memory
    tiling optimization.
    """
    def __init__(self):
        super(ModelNew, self).__init__()
        self.matmul = matmul_extension.matmul_cuda

    def forward(self, A, B):
        return self.matmul(A, B)

N = 2048 * 2

def get_inputs():
    A = torch.rand(N, N)
    B = torch.rand(N, N)
    return [A, B]

def get_init_inputs():
    return []
\end{lstlisting}
\end{tcolorbox}
\captionof{figure}{KernelBench: CUDA Matrix Multiplication -- Model Output.}
\label{fig:kernel_bench_output}

\subsection{EmbeddedC -- Example Case}
\label{sec:embedded_c_example}

\begin{tcolorbox}[title=EmbeddedC: STM32F407 PWM with TIM12 -- Problem Description, appendixbox]
\lstset{
    basicstyle=\ttfamily\footnotesize,
    breaklines=true,
    frame=none,
    columns=fullflexible,
    xleftmargin=0pt,
    xrightmargin=0pt,
}
\begin{lstlisting}
--- PROBLEM ---

STM32F407 PWM Interrupt & GPIO Alternate Function

Configure TIM12 channel 2 to output 55% duty cycle
PWM at 800 Hz on PH7 with update interrupt enabled
and TIM8_BRK_TIM12_IRQHandler.
\end{lstlisting}
\end{tcolorbox}
\captionof{figure}{EmbeddedC: STM32F407 PWM with TIM12 -- Problem Description.}
\label{fig:embedded_c_problem}

\begin{tcolorbox}[title=EmbeddedC: STM32F407 PWM with TIM12 -- Reference Code (Part 1: Infrastructure), appendixbox]
\lstset{
    basicstyle=\ttfamily\footnotesize,
    breaklines=true,
    frame=none,
    columns=fullflexible,
    xleftmargin=0pt,
    xrightmargin=0pt,
}
\begin{lstlisting}
#include <stdint.h>

// --- INFRASTRUCTURE (Always Available) ---
typedef struct {
  volatile uint32_t CR;
  volatile uint32_t PLLCFGR;
  volatile uint32_t CFGR;
  volatile uint32_t CIR;
  volatile uint32_t AHB1RSTR;
  volatile uint32_t AHB2RSTR;
  volatile uint32_t AHB3RSTR;
  uint32_t _reserved0;
  volatile uint32_t APB1RSTR;
  volatile uint32_t APB2RSTR;
  uint32_t _reserved1[2];
  volatile uint32_t AHB1ENR;
  volatile uint32_t AHB2ENR;
  volatile uint32_t AHB3ENR;
  uint32_t _reserved2;
  volatile uint32_t APB1ENR;
  volatile uint32_t APB2ENR;
} RCC_TypeDef;
#define RCC ((RCC_TypeDef *) 0x40023800)

typedef struct {
  volatile uint32_t MODER;
  volatile uint32_t OTYPER;
  volatile uint32_t OSPEEDR;
  volatile uint32_t PUPDR;
  volatile uint32_t IDR;
  volatile uint32_t ODR;
  volatile uint32_t BSRR;
  volatile uint32_t LCKR;
  volatile uint32_t AFR[2];
} GPIO_TypeDef;
#define GPIOA ((GPIO_TypeDef *) 0x40020000)
#define GPIOB ((GPIO_TypeDef *) 0x40020400)
#define GPIOC ((GPIO_TypeDef *) 0x40020800)
#define GPIOD ((GPIO_TypeDef *) 0x40020C00)
#define GPIOE ((GPIO_TypeDef *) 0x40021000)
#define GPIOH ((GPIO_TypeDef *) 0x40021C00)

typedef struct {
  volatile uint32_t ISER[8];
  uint32_t RESERVED0[24];
  volatile uint32_t ICER[8];
  uint32_t RESERVED1[24];
  volatile uint32_t ISPR[8];
  uint32_t RESERVED2[24];
  volatile uint32_t ICPR[8];
  uint32_t RESERVED3[24];
  volatile uint32_t IABR[8];
  uint32_t RESERVED4[56];
  volatile uint32_t IPR[60];
} NVIC_TypeDef;
#define NVIC ((NVIC_TypeDef *) 0xE000E100)

typedef struct {
  volatile uint32_t CR1;
  volatile uint32_t CR2;
  volatile uint32_t SMCR;
  volatile uint32_t DIER;
  volatile uint32_t SR;
  volatile uint32_t EGR;
  volatile uint32_t CCMR1;
  volatile uint32_t CCMR2;
  volatile uint32_t CCER;
  volatile uint32_t CNT;
  volatile uint32_t PSC;
  volatile uint32_t ARR;
  uint32_t _reserved1;
  volatile uint32_t CCR1;
  volatile uint32_t CCR2;
  volatile uint32_t CCR3;
  volatile uint32_t CCR4;
} TIM_TypeDef;
#define TIM12 ((TIM_TypeDef *) 0x40001800)

void TIM8_BRK_TIM12_IRQHandler(void);
volatile uint32_t tim12_update_flag = 0;
\end{lstlisting}
\end{tcolorbox}
\captionof{figure}{EmbeddedC: STM32F407 PWM with TIM12 -- Reference Code (Part 1: Infrastructure).}
\label{fig:embedded_c_infra}

\begin{tcolorbox}[title=EmbeddedC: STM32F407 PWM with TIM12 -- Reference Code (Part 2: Main \& IRQ), appendixbox]
\lstset{
    basicstyle=\ttfamily\footnotesize,
    breaklines=true,
    frame=none,
    columns=fullflexible,
    xleftmargin=0pt,
    xrightmargin=0pt,
}
\begin{lstlisting}
int main(void) {
  // 1. Enable peripheral clocks
  RCC->AHB1ENR |= (1UL << 7);   // GPIOH
  RCC->APB1ENR |= (1UL << 6);   // TIM12

  // 2. Configure PH7 as AF (TIM12_CH2)
  // PH7 mode: AF (bits 15:14 = 0b10)
  GPIOH->MODER &= ~(3UL << 14);
  GPIOH->MODER |= (2UL << 14);
  // PH7 output type: Push-pull (bit 7 = 0)
  GPIOH->OTYPER &= ~(1UL << 7);
  // PH7 speed: High (bits 15:14 = 0b11)
  GPIOH->OSPEEDR |= (3UL << 14);
  // PH7 pull: None (bits 15:14 = 0b00)
  GPIOH->PUPDR &= ~(3UL << 14);
  // PH7 AF9 (TIM12) AFR[0] bits 31:28 = 0b1001
  GPIOH->AFR[0] &= ~(0xFUL << 28);
  GPIOH->AFR[0] |= (9UL << 28);

  // 3. Configure TIM12: 800 Hz PWM, 55% duty
  // APB1 timer clock = 84 MHz
  TIM12->PSC = 0;
  // ARR = (84e6 / 800) - 1 = 104999
  TIM12->ARR = 104999UL;
  // CCMR1: OC2M=PWM mode 1, OC2PE=1
  TIM12->CCMR1 &= ~(0x7FUL << 8);
  TIM12->CCMR1 |= (0x6UL << 12) | (1UL << 11);
  // Enable CC2 output (CC2E = 1)
  TIM12->CCER |= (1UL << 4);
  // CCR2 = 0.55 * (ARR+1) = 57750
  TIM12->CCR2 = 57750UL;

  // 4. Enable update interrupt
  TIM12->DIER |= (1UL << 0); // UIE = 1

  // 5. Configure NVIC for TIM12
  // TIM8_BRK_TIM12 IRQ = 43
  // ISER[1] bit 11 (43-32=11)
  NVIC->ISER[1] |= (1UL << 11);
  NVIC->IPR[43] = 0;

  // 6. Start TIM12
  TIM12->CR1 |= (1UL << 0); // CEN = 1

  while (1) {
    for (volatile uint32_t i = 0;
         i < 10000; ++i) {
      __asm__("nop");
    }
    if (tim12_update_flag) {
      tim12_update_flag = 0;
    }
  }
}

// --- INTERRUPT HANDLER ---
void TIM8_BRK_TIM12_IRQHandler(void) {
  if (TIM12->SR & (1UL << 0)) {  // UIF
    TIM12->SR &= ~(1UL << 0);    // Clear UIF
    tim12_update_flag = 1;
  }
}
\end{lstlisting}
\end{tcolorbox}
\captionof{figure}{EmbeddedC: STM32F407 PWM with TIM12 -- Reference Code (Part 2: Main \& IRQ Handler).}
\label{fig:embedded_c_main}

\subsection{RealBench -- Example Case}
\label{sec:realbench_example}

\begin{tcolorbox}[title=RealBench: AES Cipher Top -- Design Specification, appendixbox]
\lstset{
    basicstyle=\ttfamily\footnotesize,
    breaklines=true,
    frame=none,
    columns=fullflexible,
    xleftmargin=0pt,
    xrightmargin=0pt,
}
\begin{lstlisting}
System Prompt:
You are an expert Verilog/RTL hardware design engineer.
Given a design specification, generate correct,
synthesizable Verilog code that meets all the
requirements described in the specification.

--- DESIGN SPECIFICATION ---

aes_cipher_top Design Specification

1. Introduction
The aes_cipher_top module is the core control module
of the entire AES encryption system, responsible for
coordinating and controlling the entire encryption
process. This module implements all round
transformations of the AES encryption algorithm,
including SubBytes, ShiftRows, MixColumns, and
AddRoundKey.

2. Interface

Signal     Dir     Width  Description
---------  ------  -----  ----------------------
clk        input   1      Clock signal
rst        input   1      Reset signal
ld         input   1      Load enable
done       output  1      Encryption complete
key        input   128    Input key
text_in    input   128    Input plaintext
text_out   output  128    Output ciphertext

3. Registers

Register     Width  Reset  Description
-----------  -----  -----  ----------------------
text_in_r    128    0      Input text buffer
sa[0:3][0:3] 8      0      4x4 state matrix
dcnt         4      0      Round counter (init=11)
ld_r         1      0      Load operation flag
text_out     128    0      Output ciphertext

4. Operation Principle

AES encryption performs 10 rounds of transformation
on 128-bit input data. Each round includes:
  1) SubBytes  - Non-linear byte substitution
  2) ShiftRows - Row shifting operation
  3) MixColumns - Column mixing (except final)
  4) AddRoundKey - Round key addition

State Matrix Structure:
  sa00 sa01 sa02 sa03
  sa10 sa11 sa12 sa13
  sa20 sa21 sa22 sa23
  sa30 sa31 sa32 sa33
  (4x4 byte matrix, column-major order)

ShiftRows:
  Row 0: [a b c d] -> [a b c d]  // No shift
  Row 1: [a b c d] -> [b c d a]  // Shift 1
  Row 2: [a b c d] -> [c d a b]  // Shift 2
  Row 3: [a b c d] -> [d a b c]  // Shift 3

MixColumns (in GF(2^8)):
  [02 03 01 01]   [s0]   [out0]
  [01 02 03 01] x [s1] = [out1]
  [01 01 02 03]   [s2]   [out2]
  [03 01 01 02]   [s3]   [out3]

5. State Machine

IDLE: Reset state, wait for ld=1.
INIT_ROUND: Load input, initial AddRoundKey,
  set dcnt=0xB (11).
ROUND_OP: Standard rounds 1-9.
FINAL_ROUND: Round 10 (no MixColumns).
DONE: Set done signal, hold result.

6. Submodules

aes_key_expand_128:
  Ports: clk, kld, key[127:0],
         wo_0[31:0], wo_1[31:0],
         wo_2[31:0], wo_3[31:0]

aes_sbox:
  Ports: a[7:0] (input), b[7:0] (output)

7. Constraints
  - 16 S-box modules required
  - 4x4 state matrix register array
  - One round per clock cycle
  - Register updates on rising edge
\end{lstlisting}
\end{tcolorbox}
\captionof{figure}{RealBench: AES Cipher Top -- Design Specification.}
\label{fig:realbench_spec}

\begin{tcolorbox}[title=RealBench: AES Cipher Top -- Model Output (Part 1: Declarations \& Control), appendixbox]
\lstset{
    basicstyle=\ttfamily\footnotesize,
    breaklines=true,
    frame=none,
    columns=fullflexible,
    xleftmargin=0pt,
    xrightmargin=0pt,
}
\begin{lstlisting}
module aes_cipher_top(
    clk, rst, ld, done, key, text_in, text_out);
input        clk, rst;
input        ld;
output       done;
input  [127:0] key;
input  [127:0] text_in;
output [127:0] text_out;

wire [31:0] w0, w1, w2, w3;
reg  [127:0] text_in_r;
reg  [127:0] text_out;
reg  [7:0] sa00, sa01, sa02, sa03;
reg  [7:0] sa10, sa11, sa12, sa13;
reg  [7:0] sa20, sa21, sa22, sa23;
reg  [7:0] sa30, sa31, sa32, sa33;
wire [7:0] sa00_next, sa01_next,
           sa02_next, sa03_next;
wire [7:0] sa10_next, sa11_next,
           sa12_next, sa13_next;
wire [7:0] sa20_next, sa21_next,
           sa22_next, sa23_next;
wire [7:0] sa30_next, sa31_next,
           sa32_next, sa33_next;
wire [7:0] sa00_sub, sa01_sub,
           sa02_sub, sa03_sub;
wire [7:0] sa10_sub, sa11_sub,
           sa12_sub, sa13_sub;
wire [7:0] sa20_sub, sa21_sub,
           sa22_sub, sa23_sub;
wire [7:0] sa30_sub, sa31_sub,
           sa32_sub, sa33_sub;
wire [7:0] sa00_sr, sa01_sr,
           sa02_sr, sa03_sr;
wire [7:0] sa10_sr, sa11_sr,
           sa12_sr, sa13_sr;
wire [7:0] sa20_sr, sa21_sr,
           sa22_sr, sa23_sr;
wire [7:0] sa30_sr, sa31_sr,
           sa32_sr, sa33_sr;
wire [7:0] sa00_mc, sa01_mc,
           sa02_mc, sa03_mc;
wire [7:0] sa10_mc, sa11_mc,
           sa12_mc, sa13_mc;
wire [7:0] sa20_mc, sa21_mc,
           sa22_mc, sa23_mc;
wire [7:0] sa30_mc, sa31_mc,
           sa32_mc, sa33_mc;
reg  done, ld_r;
reg  [3:0] dcnt;

// --- Round Counter ---
always @(posedge clk)
  if(!rst)    dcnt <= 4'h0;
  else if(ld) dcnt <= 4'hb;
  else if(|dcnt) dcnt <= dcnt - 4'h1;

// --- Done & Load Control ---
always @(posedge clk)
  done <= !(|dcnt[3:1]) & dcnt[0] & !ld;
always @(posedge clk)
  if(ld) text_in_r <= text_in;
always @(posedge clk)
  ld_r <= ld;
\end{lstlisting}
\end{tcolorbox}
\captionof{figure}{RealBench: AES Cipher Top -- Model Output (Part 1: Declarations \& Control).}
\label{fig:realbench_decl}

\begin{tcolorbox}[title=RealBench: AES Cipher Top -- Model Output (Part 2: State Matrix \& ShiftRows), appendixbox]
\lstset{
    basicstyle=\ttfamily\footnotesize,
    breaklines=true,
    frame=none,
    columns=fullflexible,
    xleftmargin=0pt,
    xrightmargin=0pt,
}
\begin{lstlisting}
// --- State Matrix Update (Initial Round + Next) ---
always @(posedge clk)
  sa33 <= ld_r ? text_in_r[007:000]^w3[07:00]
              : sa33_next;
always @(posedge clk)
  sa23 <= ld_r ? text_in_r[015:008]^w3[15:08]
              : sa23_next;
always @(posedge clk)
  sa13 <= ld_r ? text_in_r[023:016]^w3[23:16]
              : sa13_next;
always @(posedge clk)
  sa03 <= ld_r ? text_in_r[031:024]^w3[31:24]
              : sa03_next;
always @(posedge clk)
  sa32 <= ld_r ? text_in_r[039:032]^w2[07:00]
              : sa32_next;
always @(posedge clk)
  sa22 <= ld_r ? text_in_r[047:040]^w2[15:08]
              : sa22_next;
always @(posedge clk)
  sa12 <= ld_r ? text_in_r[055:048]^w2[23:16]
              : sa12_next;
always @(posedge clk)
  sa02 <= ld_r ? text_in_r[063:056]^w2[31:24]
              : sa02_next;
always @(posedge clk)
  sa31 <= ld_r ? text_in_r[071:064]^w1[07:00]
              : sa31_next;
always @(posedge clk)
  sa21 <= ld_r ? text_in_r[079:072]^w1[15:08]
              : sa21_next;
always @(posedge clk)
  sa11 <= ld_r ? text_in_r[087:080]^w1[23:16]
              : sa11_next;
always @(posedge clk)
  sa01 <= ld_r ? text_in_r[095:088]^w1[31:24]
              : sa01_next;
always @(posedge clk)
  sa30 <= ld_r ? text_in_r[103:096]^w0[07:00]
              : sa30_next;
always @(posedge clk)
  sa20 <= ld_r ? text_in_r[111:104]^w0[15:08]
              : sa20_next;
always @(posedge clk)
  sa10 <= ld_r ? text_in_r[119:112]^w0[23:16]
              : sa10_next;
always @(posedge clk)
  sa00 <= ld_r ? text_in_r[127:120]^w0[31:24]
              : sa00_next;

// --- ShiftRows ---
// Row 0: no shift
assign sa00_sr = sa00_sub;
assign sa01_sr = sa01_sub;
assign sa02_sr = sa02_sub;
assign sa03_sr = sa03_sub;
// Row 1: shift left 1
assign sa10_sr = sa11_sub;
assign sa11_sr = sa12_sub;
assign sa12_sr = sa13_sub;
assign sa13_sr = sa10_sub;
// Row 2: shift left 2
assign sa20_sr = sa22_sub;
assign sa21_sr = sa23_sub;
assign sa22_sr = sa20_sub;
assign sa23_sr = sa21_sub;
// Row 3: shift left 3
assign sa30_sr = sa33_sub;
assign sa31_sr = sa30_sub;
assign sa32_sr = sa31_sub;
assign sa33_sr = sa32_sub;
\end{lstlisting}
\end{tcolorbox}
\captionof{figure}{RealBench: AES Cipher Top -- Model Output (Part 2: State Matrix \& ShiftRows).}
\label{fig:realbench_state}

\begin{tcolorbox}[title=RealBench: AES Cipher Top -- Model Output (Part 3: MixColumns \& AddRoundKey), appendixbox]
\lstset{
    basicstyle=\ttfamily\footnotesize,
    breaklines=true,
    frame=none,
    columns=fullflexible,
    xleftmargin=0pt,
    xrightmargin=0pt,
}
\begin{lstlisting}
// --- MixColumns ---
wire [31:0] mix_col_out0, mix_col_out1,
            mix_col_out2, mix_col_out3;
assign mix_col_out0 =
    mix_col(sa00_sr,sa10_sr,sa20_sr,sa30_sr);
assign mix_col_out1 =
    mix_col(sa01_sr,sa11_sr,sa21_sr,sa31_sr);
assign mix_col_out2 =
    mix_col(sa02_sr,sa12_sr,sa22_sr,sa32_sr);
assign mix_col_out3 =
    mix_col(sa03_sr,sa13_sr,sa23_sr,sa33_sr);

assign {sa00_mc,sa10_mc,sa20_mc,sa30_mc}
    = mix_col_out0;
assign {sa01_mc,sa11_mc,sa21_mc,sa31_mc}
    = mix_col_out1;
assign {sa02_mc,sa12_mc,sa22_mc,sa32_mc}
    = mix_col_out2;
assign {sa03_mc,sa13_mc,sa23_mc,sa33_mc}
    = mix_col_out3;

// --- AddRoundKey (next state) ---
assign sa00_next = sa00_mc ^ w0[31:24];
assign sa01_next = sa01_mc ^ w1[31:24];
assign sa02_next = sa02_mc ^ w2[31:24];
assign sa03_next = sa03_mc ^ w3[31:24];
assign sa10_next = sa10_mc ^ w0[23:16];
assign sa11_next = sa11_mc ^ w1[23:16];
assign sa12_next = sa12_mc ^ w2[23:16];
assign sa13_next = sa13_mc ^ w3[23:16];
assign sa20_next = sa20_mc ^ w0[15:08];
assign sa21_next = sa21_mc ^ w1[15:08];
assign sa22_next = sa22_mc ^ w2[15:08];
assign sa23_next = sa23_mc ^ w3[15:08];
assign sa30_next = sa30_mc ^ w0[07:00];
assign sa31_next = sa31_mc ^ w1[07:00];
assign sa32_next = sa32_mc ^ w2[07:00];
assign sa33_next = sa33_mc ^ w3[07:00];
\end{lstlisting}
\end{tcolorbox}
\captionof{figure}{RealBench: AES Cipher Top -- Model Output (Part 3: MixColumns \& AddRoundKey).}
\label{fig:realbench_mix}

\begin{tcolorbox}[title=RealBench: AES Cipher Top -- Model Output (Part 4: Output \& Submodules), appendixbox]
\lstset{
    basicstyle=\ttfamily\footnotesize,
    breaklines=true,
    frame=none,
    columns=fullflexible,
    xleftmargin=0pt,
    xrightmargin=0pt,
}
\begin{lstlisting}
// --- Output Generation (Final Round) ---
always @(posedge clk)
  text_out[127:120] <= sa00_sr ^ w0[31:24];
always @(posedge clk)
  text_out[095:088] <= sa01_sr ^ w1[31:24];
always @(posedge clk)
  text_out[063:056] <= sa02_sr ^ w2[31:24];
always @(posedge clk)
  text_out[031:024] <= sa03_sr ^ w3[31:24];
always @(posedge clk)
  text_out[119:112] <= sa10_sr ^ w0[23:16];
always @(posedge clk)
  text_out[087:080] <= sa11_sr ^ w1[23:16];
always @(posedge clk)
  text_out[055:048] <= sa12_sr ^ w2[23:16];
always @(posedge clk)
  text_out[023:016] <= sa13_sr ^ w3[23:16];
always @(posedge clk)
  text_out[111:104] <= sa20_sr ^ w0[15:08];
always @(posedge clk)
  text_out[079:072] <= sa21_sr ^ w1[15:08];
always @(posedge clk)
  text_out[047:040] <= sa22_sr ^ w2[15:08];
always @(posedge clk)
  text_out[015:008] <= sa23_sr ^ w3[15:08];
always @(posedge clk)
  text_out[103:096] <= sa30_sr ^ w0[07:00];
always @(posedge clk)
  text_out[071:064] <= sa31_sr ^ w1[07:00];
always @(posedge clk)
  text_out[039:032] <= sa32_sr ^ w2[07:00];
always @(posedge clk)
  text_out[007:000] <= sa33_sr ^ w3[07:00];

// --- GF(2^8) Functions ---
function [7:0] xtime;
  input [7:0] b;
  xtime = {b[6:0],1'b0}
        ^ (8'h1b & {8{b[7]}});
endfunction

function [31:0] mix_col;
  input [7:0] s0, s1, s2, s3;
  reg [7:0] s0_o, s1_o, s2_o, s3_o;
  begin
    mix_col[31:24] =
        xtime(s0)^xtime(s1)^s1^s2^s3;
    mix_col[23:16] =
        s0^xtime(s1)^xtime(s2)^s2^s3;
    mix_col[15:08] =
        s0^s1^xtime(s2)^xtime(s3)^s3;
    mix_col[07:00] =
        xtime(s0)^s0^s1^s2^xtime(s3);
  end
endfunction

// --- Key Expansion Submodule ---
aes_key_expand_128 u0(
    .clk(clk), .kld(ld), .key(key),
    .wo_0(w0), .wo_1(w1),
    .wo_2(w2), .wo_3(w3));

// --- 16 S-Box Instances ---
aes_sbox us00(.a(sa00), .b(sa00_sub));
aes_sbox us01(.a(sa01), .b(sa01_sub));
aes_sbox us02(.a(sa02), .b(sa02_sub));
aes_sbox us03(.a(sa03), .b(sa03_sub));
aes_sbox us10(.a(sa10), .b(sa10_sub));
aes_sbox us11(.a(sa11), .b(sa11_sub));
aes_sbox us12(.a(sa12), .b(sa12_sub));
aes_sbox us13(.a(sa13), .b(sa13_sub));
aes_sbox us20(.a(sa20), .b(sa20_sub));
aes_sbox us21(.a(sa21), .b(sa21_sub));
aes_sbox us22(.a(sa22), .b(sa22_sub));
aes_sbox us23(.a(sa23), .b(sa23_sub));
aes_sbox us30(.a(sa30), .b(sa30_sub));
aes_sbox us31(.a(sa31), .b(sa31_sub));
aes_sbox us32(.a(sa32), .b(sa32_sub));
aes_sbox us33(.a(sa33), .b(sa33_sub));

endmodule
\end{lstlisting}
\end{tcolorbox}
\captionof{figure}{RealBench: AES Cipher Top -- Model Output (Part 4: Output Generation \& Submodules).}
\label{fig:realbench_output}

\subsection{ArchXBench -- Example Case}
\label{sec:archxbench_example}

\begin{tcolorbox}[title=ArchXBench: 8-to-3 Binary Encoder -- Design Specification, appendixbox]
\lstset{
    basicstyle=\ttfamily\footnotesize,
    breaklines=true,
    frame=none,
    columns=fullflexible,
    xleftmargin=0pt,
    xrightmargin=0pt,
}
\begin{lstlisting}
System Prompt:
You are an expert Verilog/RTL hardware design engineer.
Given a design specification, generate correct,
synthesizable Verilog code that meets all the
requirements described in the specification.

--- DESIGN SPECIFICATION ---

Title: 8-to-3 Binary Encoder

Objective:
Design an 8-to-3 binary encoder that outputs the binary
representation of the index of the highest active input,
assuming one-hot input.

Background:
Binary encoders are combinational circuits that reduce
multiple input lines to a binary representation. A binary
encoder assumes that only one input is active (one-hot
input), suitable for controlled environments where inputs
are guaranteed to be mutually exclusive.

Design Constraints:
- Assumes only one input is high at a time.
- Purely combinational design.
- When no bits set or more than one bit set, drive
  outputs to 0.
- Any X or Z on any input bit must propagate through
  so that corresponding output bits become X.

Design Name: binary_encoder_8to3

Inputs:
  in[7:0]   // One-hot 8-bit input

Outputs:
  out[2:0]  // 3-bit binary output

Module Signature:
  module binary_encoder_8to3 (
      input  [7:0] in,
      output [2:0] out
  );

Design Notes:
- Assumes exactly one input is high.
- Outputs the corresponding binary index.
- Use purely combinational logic.
\end{lstlisting}
\end{tcolorbox}
\captionof{figure}{ArchXBench: 8-to-3 Binary Encoder -- Design Specification.}
\label{fig:archxbench_spec}

\begin{tcolorbox}[title=ArchXBench: 8-to-3 Binary Encoder -- Model Output, appendixbox]
\lstset{
    basicstyle=\ttfamily\footnotesize,
    breaklines=true,
    frame=none,
    columns=fullflexible,
    xleftmargin=0pt,
    xrightmargin=0pt,
}
\begin{lstlisting}
module binary_encoder_8to3 (
    input  [7:0] in,
    output [2:0] out
);

  // Check for any X or Z in input
  wire has_xz;
  assign has_xz = (^in === 1'bx);

  // Count number of 1's in input
  wire [3:0] count_ones;
  assign count_ones =
      {1'b0, in[0]} + {1'b0, in[1]} +
      {1'b0, in[2]} + {1'b0, in[3]} +
      {1'b0, in[4]} + {1'b0, in[5]} +
      {1'b0, in[6]} + {1'b0, in[7]};

  // Check if exactly one bit is high
  wire valid_one_hot;
  assign valid_one_hot =
      (count_ones == 4'd1);

  // Encoder for position (one-hot input)
  wire [2:0] encoded_out;
  assign encoded_out[2] =
      in[7] | in[6] | in[5] | in[4];
  assign encoded_out[1] =
      in[7] | in[6] | in[3] | in[2];
  assign encoded_out[0] =
      in[7] | in[5] | in[3] | in[1];

  // Final output with X propagation
  // and validity check
  assign out = has_xz ? 3'bxxx :
      (valid_one_hot ? encoded_out : 3'b000);

endmodule

--- ALTERNATIVE IMPLEMENTATION (casez) ---

module binary_encoder_8to3 (
    input  [7:0] in,
    output [2:0] out
);

  wire has_xz;
  assign has_xz = (^in === 1'bx);

  reg [3:0] count;
  integer i;
  always @(*) begin
    count = 4'd0;
    for (i = 0; i < 8; i = i + 1) begin
      if (in[i] === 1'b1)
        count = count + 1;
    end
  end

  wire valid_one_hot;
  assign valid_one_hot = (count == 4'd1);

  reg [2:0] encoded_out;
  always @(*) begin
    encoded_out = 3'b000;
    casez (in)
      8'b00000001: encoded_out = 3'b000;
      8'b00000010: encoded_out = 3'b001;
      8'b00000100: encoded_out = 3'b010;
      8'b00001000: encoded_out = 3'b011;
      8'b00010000: encoded_out = 3'b100;
      8'b00100000: encoded_out = 3'b101;
      8'b01000000: encoded_out = 3'b110;
      8'b10000000: encoded_out = 3'b111;
      default:     encoded_out = 3'b000;
    endcase
  end

  assign out = has_xz ? 3'bxxx :
      (valid_one_hot ? encoded_out : 3'b000);

endmodule

--- KEY DESIGN DECISIONS ---

1. X/Z propagation:
   Using === 1'bx checks for X or Z values.
   If any input bit is X/Z, the reduction XOR
   returns X.

2. Valid one-hot detection:
   Count the number of 1's in the input. If
   exactly one bit is high, it's valid.

3. Output selection:
   - X/Z detected    -> output 3'bxxx
   - Valid one-hot   -> output encoded position
   - Otherwise       -> output 3'b000


\end{lstlisting}
\end{tcolorbox}
\captionof{figure}{ArchXBench: 8-to-3 Binary Encoder -- Model Output.}
\label{fig:archxbench_output}

\end{document}